\documentclass[12pt,a4paper,final,appendixprefix,bibtotoc]{scrbook}

\usepackage{ifpdf}
\ifpdf
\pdfoutput=1
\fi
\usepackage[latin1]{inputenc}
\usepackage{url}
\usepackage{wrapfig}
\usepackage{graphicx}
\usepackage{color}
\usepackage{colortbl}
\usepackage{ifthen}
\usepackage{supertabular}
\usepackage{pdflscape}
\usepackage[absolute]{textpos}
\usepackage{subfigure}
\usepackage{setspace}
\usepackage{graphics}
\usepackage{amssymb}
\usepackage{amsmath}
\usepackage{epsfig}
\usepackage{listings}
\usepackage{times}

\renewcommand{\subfigtopskip}{0pt}
\renewcommand{\subfigbottomskip}{0pt}

\definecolor{mygrey}{rgb}{0.45,0.45,0.45}
\definecolor{mydarkgrey}{rgb}{0.2,0.2,0.2}
\definecolor{red}{rgb}{1.0,0.33,0.33}
\definecolor{orange}{rgb}{1.00,0.73,0.33}
\definecolor{yellow}{rgb}{0.95,0.92,0.}
\definecolor{lightgreen}{rgb}{0.3,0.95,0.46}
\definecolor{titleblue}{rgb}{0.03,0.10,0.46}

\usepackage{typearea}\areaset[1.5cm]{418pt}{658pt}
\author{}
\title{}

\begin{document}

\setcounter{tocdepth}{3}
\setcounter{secnumdepth}{3}
\frontmatter

\begin{titlepage}
\vspace*{-1cm}
\newlength{\links}
\setlength{\links}{0.9cm}
\setlength{\TPHorizModule}{1cm}
\setlength{\TPVertModule}{1cm}
\textblockorigin{0pt}{0pt}

\sf
\LARGE

\begin{textblock}{16.5}(2.8,2.6)
 \hspace*{-0.65cm} \textbf{UNIVERSITÄT DUISBURG-ESSEN} \\
 \hspace*{-1.2cm} \rule{5mm}{5mm} \hspace*{0.05cm} FAKULTÄT FÜR INGENIEURWISSENSCHAFTEN\\INTERNATIONAL STUDIES IN ENGINEERING\\MAJOR COMPUTER ENGINEERING\\
\end{textblock}

\begin{textblock}{14.5}(2.8,8.5)
  \large
\noindent {\bf Master Thesis} \\[1cm]
{\LARGE \Large\bf Simulation of Resource Usage in Parallel Evolutionary Peptide Optimization using JavaSpaces Technology} \\[1.5cm]
Andias Tektoniko Wira Alam\\
Matrikelnummer: 2218260
\end{textblock}

\begin{textblock}{7.6}(11.5,17.5)
\includegraphics[scale=0.88]{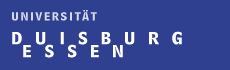}\\
\normalsize
\raggedleft

Zentrum für Medizinische Biotechnologie \\

Abteilung Bioinformatik \\
Universität Duisburg-Essen \\[2ex]

January 30, 2008\\[15ex]
\raggedright

{\bf Supervisor:} \\
Prof. Dr. Daniel Hoffmann\\

\end{textblock}

\end{titlepage}

\chapter*{Acknowledgements}

This thesis has my name on it, but there are many people who contribute
to its completion. I would like to acknowledge the following people:

\doublespacing
\singlespacing
My thesis supervisor, Prof. Dr. Daniel Hoffmann, he opened me the gate
to the scientific world. I am indebted to him for the guidance, advice,
motivation and kindness during I worked with him.

\doublespacing
\singlespacing
My thesis advisor, Manuel Prinz, he is a really nice person. He has
almost the same hobbies as mine, Linux and Heavy Metal. I really thank
for his advice, encouragement, enduring patience and constant support.

\doublespacing
\singlespacing
My lecturer, Prof. Dr. Heinz Ulrich Hoppe, he has introduced JavaSpaces
Technology in his great lecture of Distributed Systems. 

\doublespacing
\singlespacing
My colleagues at the Department of Bioinformatics, 
Julia Goldmann, Oliver Kuhn, Björn Thorwirth and Stanislav Jakuschev,
they are all good friends.

\doublespacing
\singlespacing
My wife, Ika Yuli Rakhmawati, she is the special one in my life. 
You are the greatest partner in the world who really understands me a lot.
I love you, I owe you one.

\doublespacing
\singlespacing
My family members who always support me by phone and SMS, I am honored and
grateful to be a member in this family.

\doublespacing
\singlespacing
My thesis relied heavily on a number of free software packages, and their 
authors have my deeply appreciation.

\tableofcontents

\chapter*{Abstract}
Peptide Optimization is a highly complex problem and it takes very 
long time of computation. This optimization process uses many software 
applications in a cluster running GNU/Linux Operating System that perform 
special tasks. The application to organize the whole optimization process 
had been already developed, namely SEPP (System for Evolutionary Pareto 
Optimization of Peptides/Polymers). A single peptide optimization takes
a lot of computation time to produce a certain number of individuals. However, 
it can be accelerated by increasing the degree of parallelism as well as 
the number of nodes (processors) in the cluster.

\doublespacing
\singlespacing
\noindent
In this master thesis, I build a model simulating the interplay of the programs 
so that the usage of each resource (processor) can be determined and also the 
approximated time needed for the overall optimization process. There are two 
Evolutionary Algorithms that could be used in the optimization, 
namely Generation-based and 
Steady-state Evolutionary Algorithm. The results of each Evolutionary Algorithm 
are shown based on the simulations. Moreover, the results are also compared by 
using different parameters (the degree of parallelism and the number of 
processors) in the simulation to give an overview of the advantages and the 
disadvantages of the algorithms in terms of computation time and resource usage. 
The model is built up using JavaSpaces Technology.

\doublespacing
\singlespacing
\noindent
{\it Keywords:} Multiobjective optimization; Evolutionary Algorithms; 
Parallel simulation; Timed Petri nets; JavaSpaces Technology.

\mainmatter

\chapter{Introduction}

\section{Peptide}

\doublespacing

\begin{figure}[htp]
\centering
\includegraphics[scale=0.5]{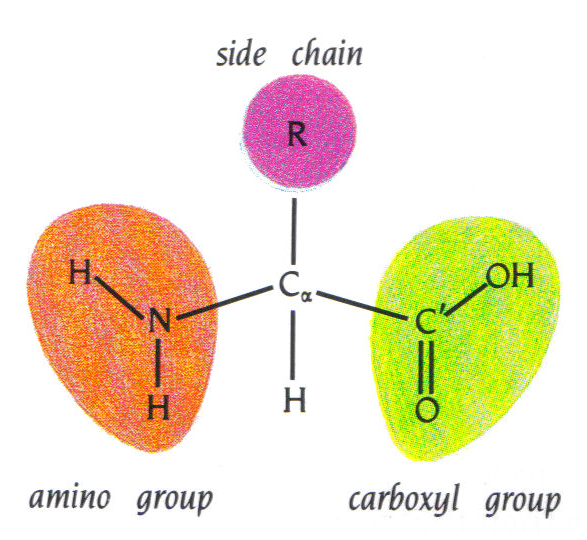}
\caption{A schematic diagram of an amino acid \cite{Branden1991}}
\label{Peptide01}
\end{figure}

\begin{figure}[htp]
\centering
\includegraphics[scale=0.4]{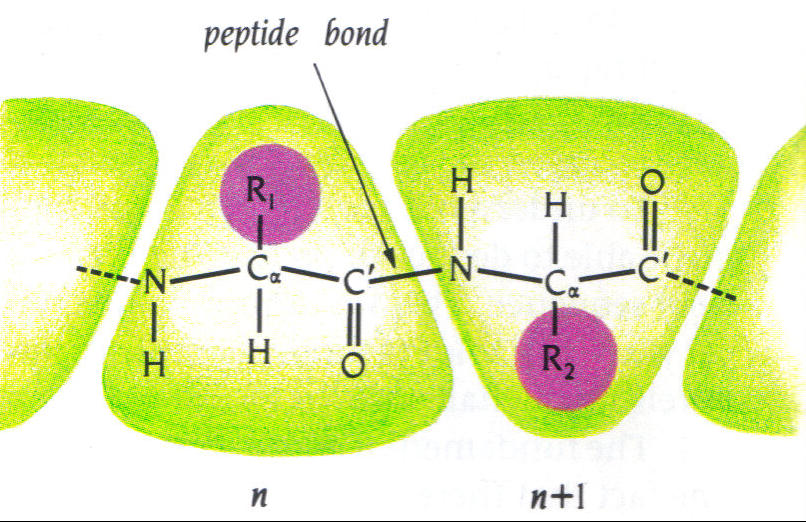}
\caption{A polypeptide chain and a peptide bond \cite{Branden1991}}
\label{Peptide02}
\end{figure}

A peptide is a polymer consisting of amino acids which are linked together. 
A schematic diagram of an amino acid is depicted in Figure \ref{Peptide01}.
A central carbon atom $\mathrm{C}_{\alpha}$ is attached to an amino group,
$\mathrm{NH}_2$, a carboxy group COOH, a hydrogen atom, H, and a side
chain, R. In a polypeptide chain the carboxy group of an amino acid $n$ 
forms a peptide bond, C-N, to the amino group of amino acid $n+1$, as shown
in Figure \ref{Peptide02} \cite{Branden1991}.

Peptide units (sequence of amino acids) are building blocks of protein structures. 
There are 20 different amino acids 
grouped in three different categories: {\it hydrophobic}, {\it charged},
and {\it polar} amino acids \cite{Branden1991}.

\section{Evolutionary Algorithm}

In this section, a short general description of the Evolutionary Algorithm
is addressed. This algorithm covers an evolutionary computation
and a theoretical framework of multicriteria decision making.

\subsection{Multiobjective Optimization}

In multiobjective optimization, there are several possibly contradicting
objectives to be optimized simultaneously. In general, a single optimal solution 
does no longer exist, but instead a set of possible solutions of
equivalent quality which are all optimal in some sense. 
Note that to obtain the optimal solution, there
is a set of optimal trade-offs between the contradicting objectives.

According to \cite{Coello1999,Taboada2007}, a multiobjective optimization problem can 
be written in the form: 

\begin{equation}
\centering
\mathrm{minimize/maximize} [f_1(x), f_2(x),...,f_k(x)]
\end{equation}

\noindent
subject to the $m$ inequality constraints:

\begin{equation}
\centering
g_i(x) \geq 0 ~ \forall i \in \{1,2,...,m\}
\label{inequalityconstraint}
\end{equation}

\noindent
and the $p$ equality constraints:

\begin{equation}
\centering
h_i(x) = 0 ~ \forall i \in \{1,2,...,p\}
\label{equalityconstraint}
\end{equation}

\noindent
where $k$ is the number of objective functions $f_i:\Re^n \rightarrow \Re$.
The vector of decision variables is $x=[x_1,x_2,...,x_n]^T$.
The aim is to determine the particular set of values 
$[x_1^*, x_2^*,...,x_n^*]$ which yield the optimum values for all 
the objective functions from among the set $F$ of all vectors 
satisfying Equation \ref{inequalityconstraint} and \ref{equalityconstraint}.

\subsection{Pareto-optimal Solutions}

In a multiobjective optimization, a solution could be best, worst, and also
indifferent to the other solutions (neither dominating or dominated) with
respect to the objective values. Consequently, there is no unique solution
to multiobjective optimization problems, but instead, there are a number
of feasible solutions available. 

\begin{figure}[htp]
  \begin{center}
     \subfigure[for a minimization problem]{\label{paretomin}\includegraphics[scale=1]{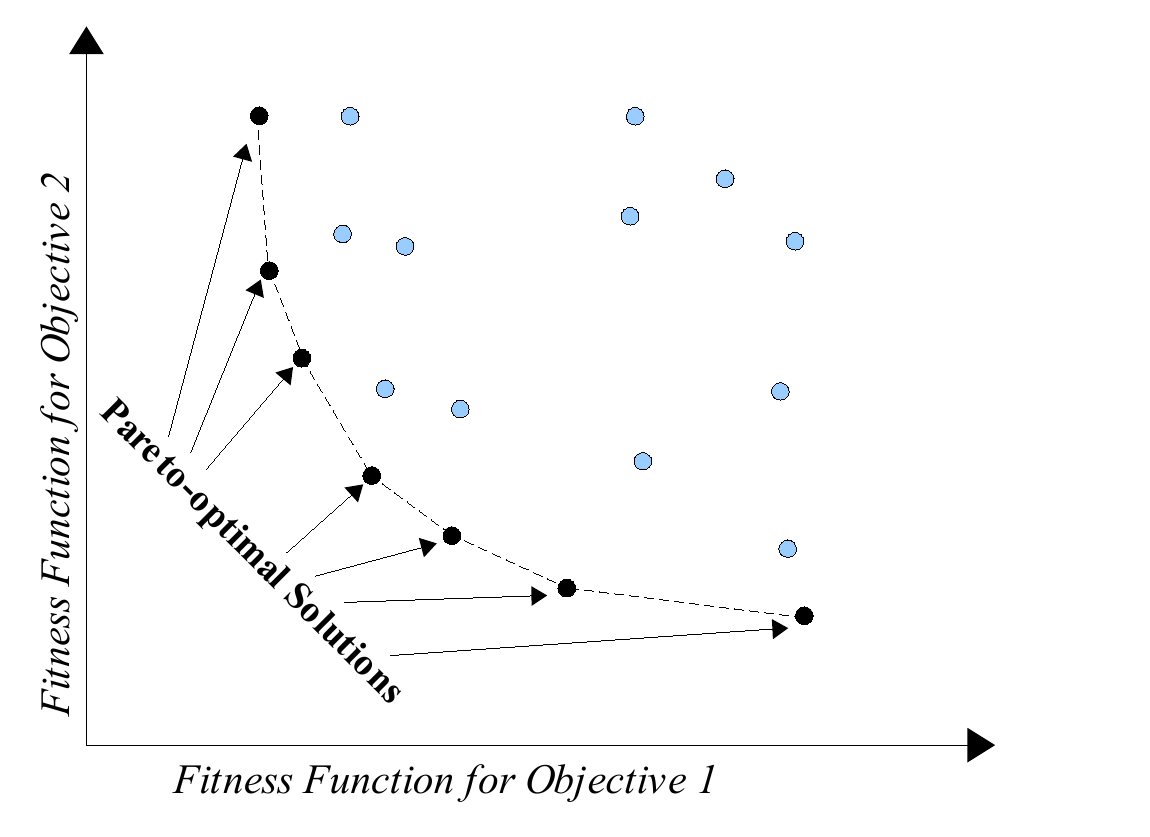}}
     \subfigure[for a maximization problem]{\label{paretomax}\includegraphics[scale=1]{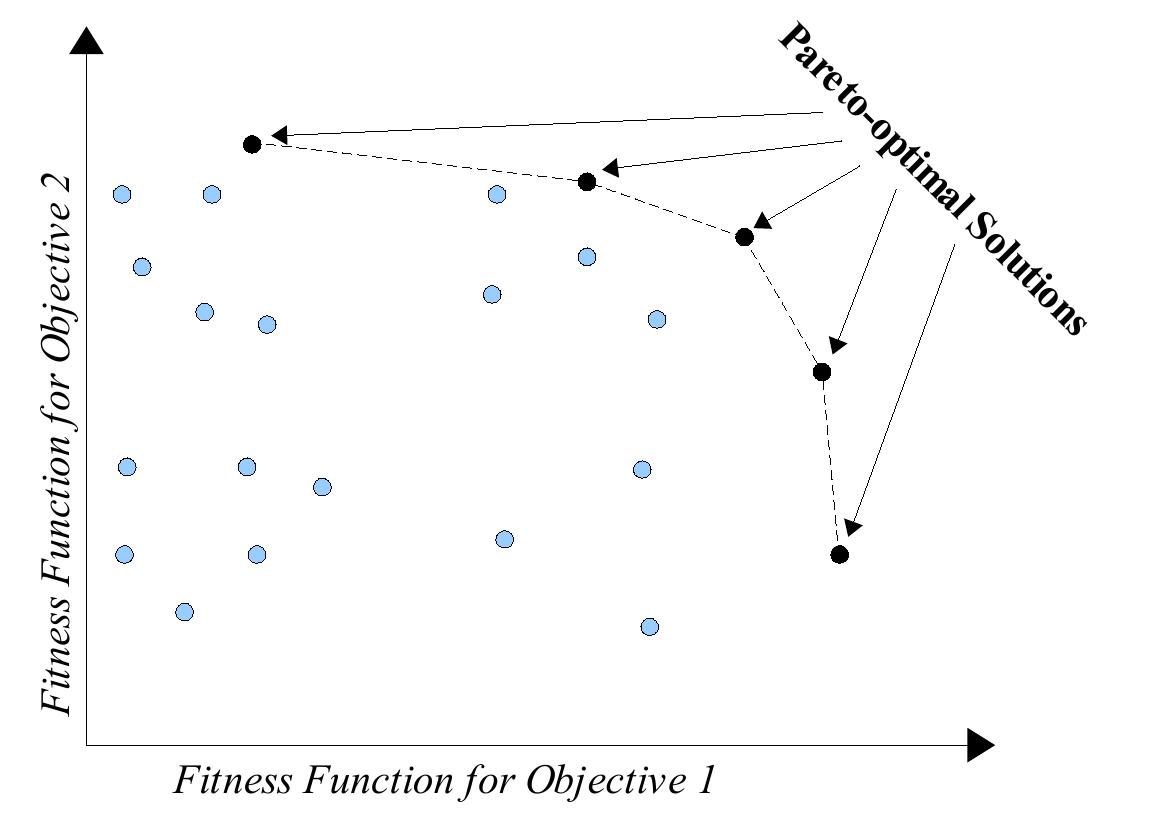}}
  \end{center}
\caption{Pareto-optimal solutions \cite{Abraham2005}}
\label{ParetoOptimalSolutions}
\end{figure}

An optimal solution is the solution that is not dominated by any other solution
in the search space. In addition, best solution means a solution which is
not worst in any of the objectives and at least better in one objective than
the other solutions in the search space \cite{Baeck1996}.
Such an optimal solution is called Pareto-optimal, whereas
the entire set of such optimal trade-off solutions is called Pareto-optimal
set. Figure \ref{ParetoOptimalSolutions} shows possible solutions in the 
search space\footnote{for the sake of simplicity, this space only represents 
two-dimensional objectives, but it's also analogous for $n$-dimensional objectives.} 
and the Pareto-optimal solutions \cite{Abraham2005}.
More precisely, for a minimization problem, a vector of decision variables 
$x^* \in F$ is called Pareto-optimal
if  $\nexists x \in F$ such that $f_i(x) \leq f_i(x^*) ~ \forall i=\{1,2,...,k\}$,
with at least one strict inequality: $f_j(x) < f_j(x^*)$ for at least one $j$.
Similarly, for a maximization problem, $x^* \in F$ is called Pareto-optimal
if  $\nexists x \in F$ such that $f_i(x) \geq f_i(x^*) ~ \forall i=\{1,2,...,k\}$,
with at least one strict inequality: $f_j(x) > f_j(x^*)$ for at least one $j$.

\subsection{Evolutionary Algorithm for Multiobjective Optimization}

\begin{figure}[htp]
\centering
\includegraphics[scale=0.8]{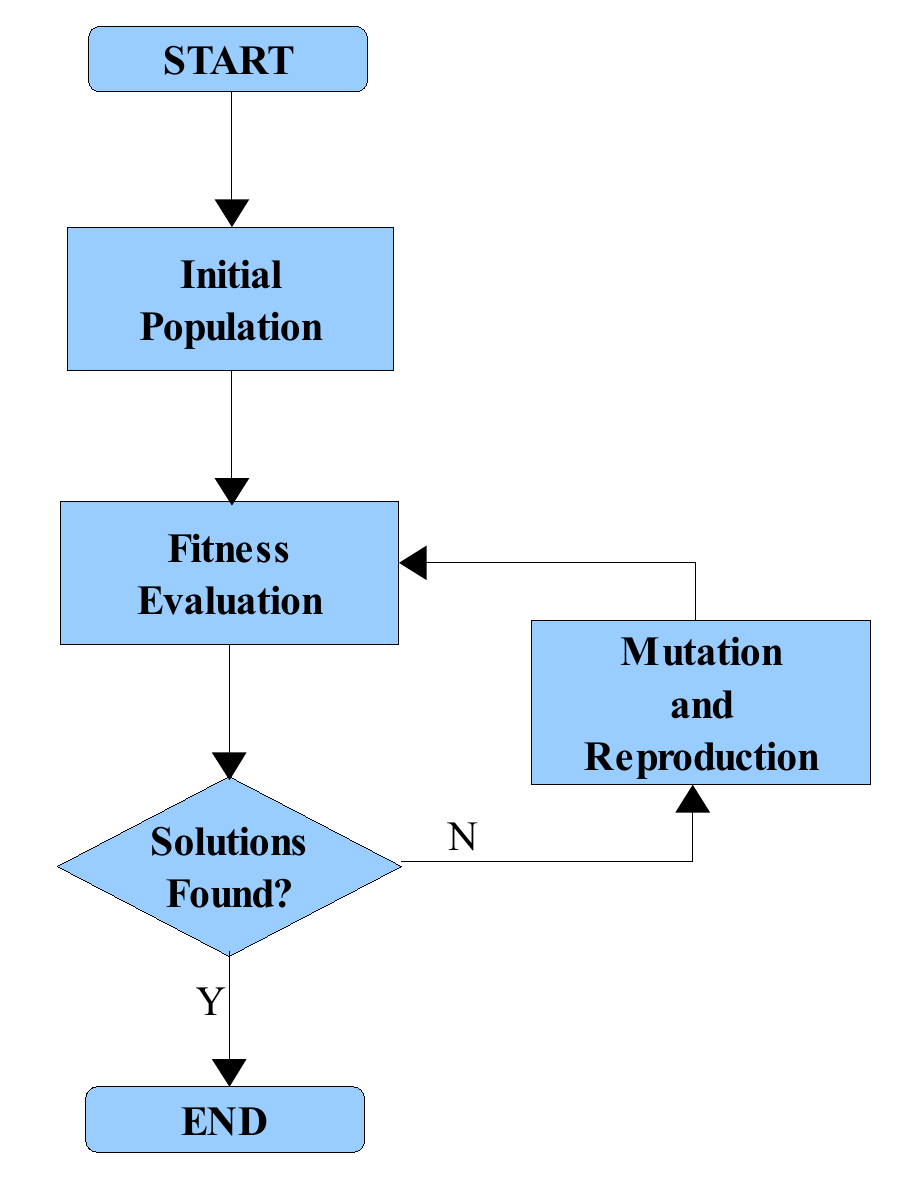}
\caption{The iterative computation process of an evolutionary algorithm \cite{Abraham2005}}
\label{EvolutionaryAlgorithmIteration}
\end{figure}

Evolutionary algorithm is characterized by a population of existing candidates.
The mutation and reproduction process enable the combination of existing solutions
to generate new solutions. 
The general iterative computation process of an evolutionary algorithm is illustrated 
in Figure \ref{EvolutionaryAlgorithmIteration} \cite{Abraham2005}.

In the population, there are a number of individuals. A natural selection by fitness functions
determines which individuals of the current population participate in the new population.
After selection, a set of potential solutions, which are all optimal in some sense, is 
eventually produced by the algorithm. But notice that these solutions are not a final decision.
There is a possibility to find better solutions by mutating the individuals. The mutation
process makes replicas or copies of the individuals, in such a way that
the characteristics (here, the sequence) of each individual are varied (mutated) based
on stochastic processes. One of the advantages is that many types of solutions are possible
to achieve. Some of the other advantages of using evolutionary algorithms is that they can be 
implemented in a parallel environment.

\section{Evolutionary Peptide Optimization}

A peptide (or protein) has a biological function. The biological function of 
a peptide depends on its three-dimensional structure characterized by its amino 
acid sequence. 
It necessarily follows that the amino acid sequence plays a major role
to ensure the biological function of peptides.
The things become interesting because the biological function of peptides
is certainly possible to be optimized. In other words, optimizing a 
biological function of a peptide means finding its optimal sequence of 
amino acids. Note that this optimization is not a single-objective 
but a multiobjective optimization, such as conformational stability,
peptide length, degree of similarity, etc.

\begin{figure}[htp]
\centering
\includegraphics[scale=1]{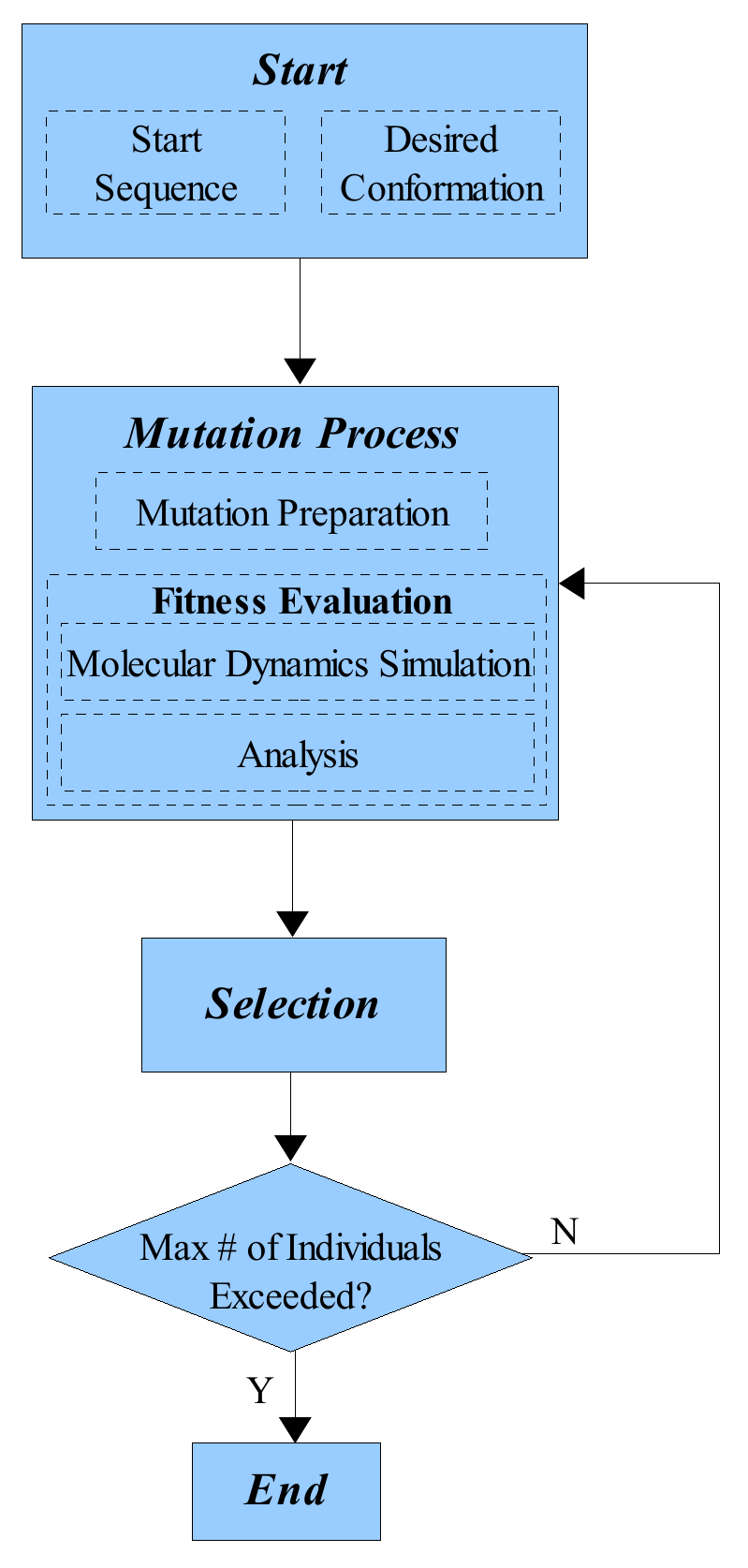}
\caption{A simplified scheme of the Evolutionary Algorithm used
for Peptide Optimization}
\label{PeptideOptimization}
\end{figure}

There is a quasi-natural approach for the finding of optimal sequence of amino
acids. This approach is based on Evolutionary Algorithms: Generation-based 
and Steady-state Algorithm \cite{Hohm2005,Prinz2006}. 
The algorithms change given peptide sequences towards sequences with 
increased propensity for a specific conformation. 
Figure \ref{PeptideOptimization} depicts a simplified scheme of 
the Evolutionary Algorithm used for Peptide Optimization.
In fact, the mutation process is able to run in parallel environment, which means
that multiple mutation processes can run concurrently. The explanation about
this concurrency in technical sense is addressed in Chapter \ref{chapter3}.

The optimization process stops after a predefined number of individuals exceeded.
In the mutation process there are four steps of computation which are
executed sequentially: compute mutation site probabilities, choose site, 
compute amino acid exchange probabilities and choose amino acid \cite{Hohm2005}.

\subsection{Generation-based and Steady-state Evolutionary Algorithm}

As previously indicated, there are two types of algorithms being 
implemented in peptide optimization, namely Generation-based and 
Steady-state Evolutionary Algorithm. The main difference is in the occurrences of
the selection process \cite{Prinz2006}. 
The Generation-based Evolutionary Algorithm performs selection
after the mutation processes (including the fitness evaluation) have been
done for all individuals. The Steady-state Evolutionary Algorithm apparently
performs selections right after one individual has been mutated. 

\begin{figure}[htp]
\centering
\includegraphics[scale=1]{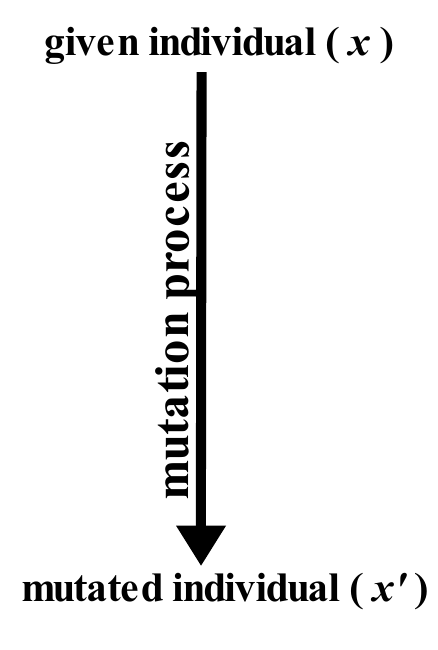}
\caption{A mutation process of an individual}
\label{MutationProcess01}
\end{figure}

\begin{figure}[htp]
\centering
\includegraphics[scale=1]{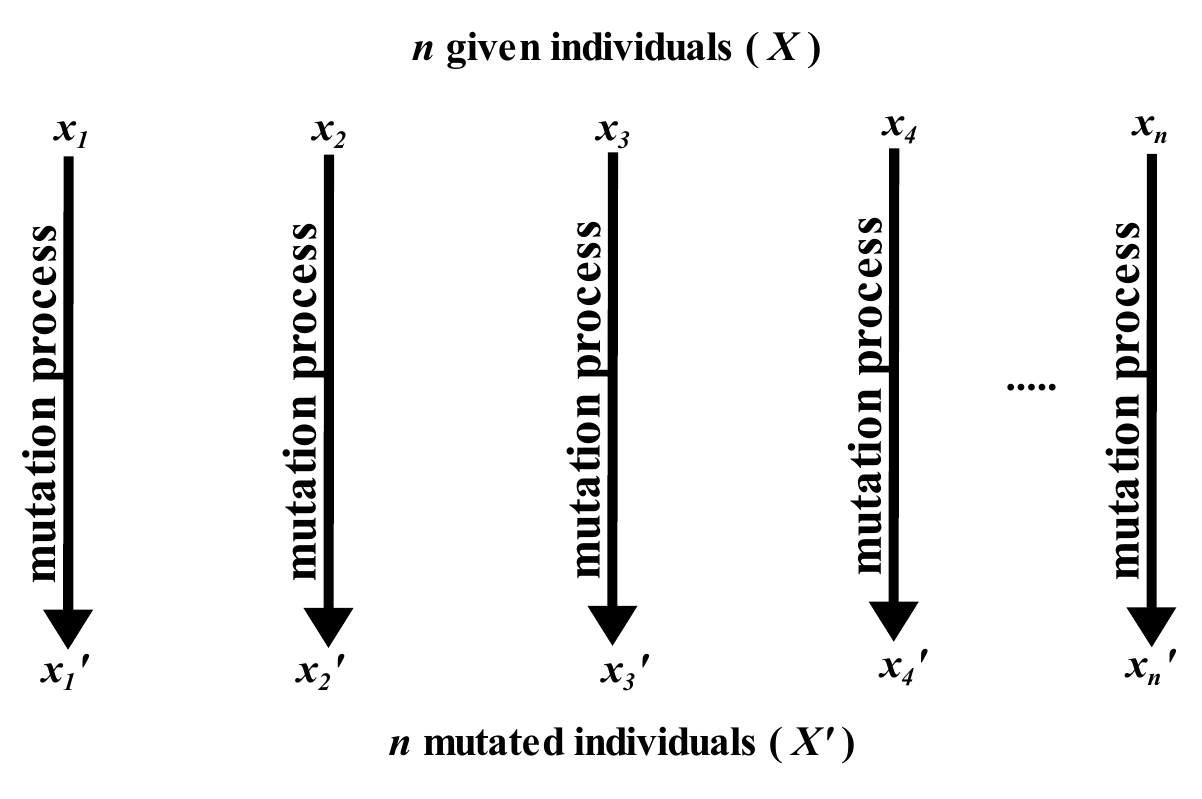}
\caption{A simultaneous mutation process of $n$ individuals}
\label{MutationProcess02}
\end{figure}

A mutation process of an individual is depicted in Figure
\ref{MutationProcess01}. Note that the computation time of a mutation process 
ranges in a set of values and can not be exactly predicted.
In parallel environment, a simultaneous mutation process of $n$ individuals
is possible. In other words, a simultaneous mutation process of $n$ individuals
consists of $n$ mutation processes running concurrently.
Figure \ref{MutationProcess02} illustrates this kind of mutation
process. 

\begin{figure}[htp]
\centering
\includegraphics[scale=1]{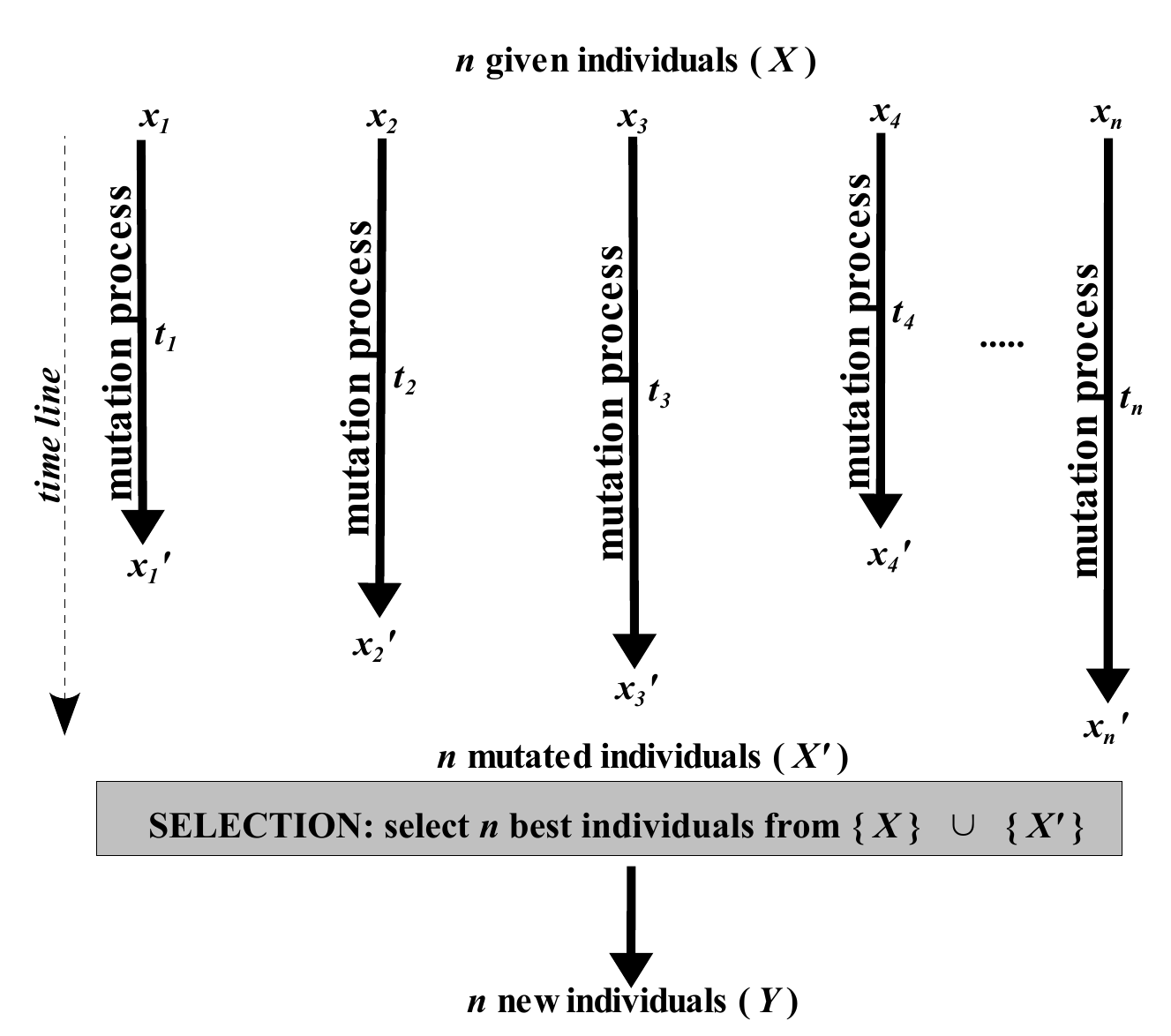}
\caption{A selection of the Generation-based Algorithm}
\label{MutationProcess03}
\end{figure}

The drawback of using Generation-based Evolutionary Algorithm is in the 
overall computation time. Every selection process depends on the longest 
mutation process among individuals. This kind of selection is illustrated
in Figure \ref{MutationProcess03}. Notice that a mutated individual doesn't necessarily
become a new individual. A new individual means an individual produced after
a selection process.
Consider that $n$ individuals being computed using $n$ processors 
and one individual corresponds to one processor, then at most
$n-1$ processors have idle state which supposedly could be used for other computations.
Moreover, there will be always $2n$ individuals in the selection process.
So if the simultaneous mutation and selection process is repeated 
so many times, hence the idle state of processors becomes higher.

In contrast, the Steady-state Evolutionary Algorithm has an advantage that the idle
state of each processor can be reduced or even omitted, because
each processor does not need to wait until all individuals have been
mutated \cite{Prinz2006}. Nevertheless, the amount of the
individuals in the selection process is consequently reduced, which is 
always only $n+1$ individuals.

\begin{figure}[htp]
\centering
\includegraphics[scale=1]{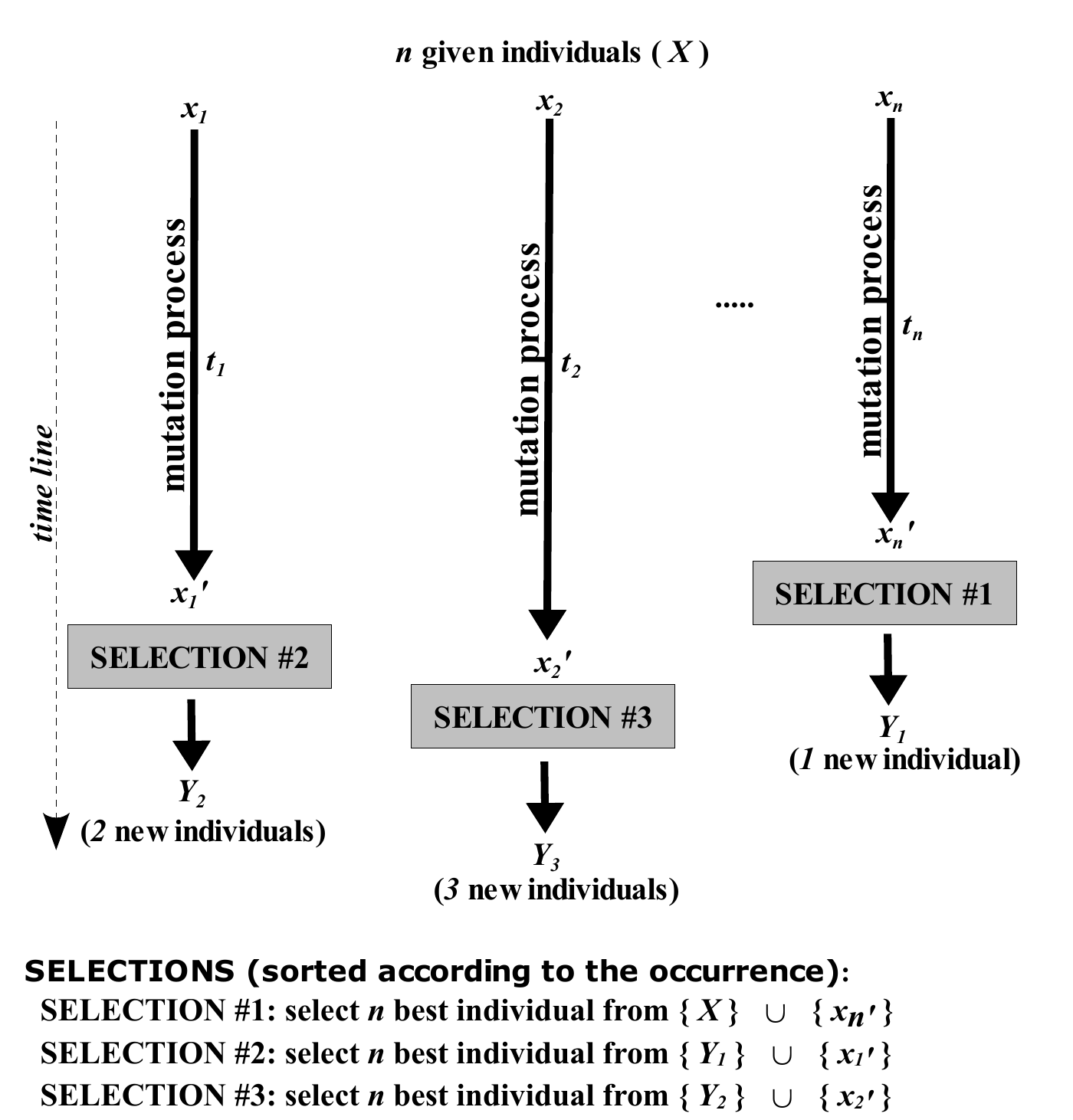}
\caption{A selection of the Steady-state Algorithm}
\label{MutationProcess04}
\end{figure}

The selection process of the Steady-state Algorithm is depicted 
in Figure \ref{MutationProcess04}. In the selection process, the best individual
replaces the worst one. Hence, the new population always consists of $n$ individuals.
In other words, the selection process selects $n$ best individuals from $n+1$ individuals,
which are $n$ from the old population and $1$ mutated individuals.
In such an unlikely case that two or several 
mutation processes finish exactly at the same time, there are two options to 
execute the selection process: simultaneously or in a random sequence. The first
option causes the selection processes have the same previous (old) population and
produces several populations. The union of those populations may have more than $n$ 
individuals which are not expected because of consistency.
Instead, in the second option, each selection has probably different old population
and the new population is always consistent of having $n$ individuals.
Notice that the execution time of the selection process, in fact, doesn't take
so much time as the mutation process.

Let's consider if the simultaneous mutation and selection process 
is repeated $k$ times and $\vec{t}_i$ is a vector contains times of all mutation
processes at $i$-th simultaneous mutation process. It follows that the overall
computation time using the Generation-based Algorithm is 
$\sum_{i=1}^{k} max(\vec{t}_i)$, whereas the Steady-state Algorithm is
$\sum_{i=1}^{k} \mu(\vec{t}_i)$. Which algorithm yields the better results 
(quality of individuals) in terms of biological function still has 
to be examined and it is not covered in this thesis.

\newpage
\section{Problem Description}

In this master thesis, the following problems are addressed, such as

\noindent
{\bf (a) Developing a Simulation Program}

There is a need for having such a program which can simulate the resource usage
in parallel evolutionary peptide optimization. The program models 
the real situation that the optimization is done on a cluster using
multiple resources (processors) and based on two evolutionary algorithms: 
Generation-based and Steady-state Algorithm. Moreover, the number of individuals 
(input and output), the degree of parallelism, and the number of resources (processors)
are variable.

\noindent
{\bf (b) Approximation of Overall Computation Time}

There is a need to estimate the overall computation time for producing
$N$ new individuals given $\eta$ individuals, where $N > \eta$.
Under this condition, the simulation runs using two different algorithms: 
Generation-based and Steady-state Algorithm. Besides, in the simulation some 
parameters are varied, such as the number of new individuals ($N$) and the number 
of resources. 
The results are then analyzed in order to evaluate the efficiency of the algorithms
with respect to the number of new individuals and the overall computation time.

\noindent
{\bf (c) Efficiency of Resource Usage}

There is a need to determine the efficiency of the resource usage. Comparing 
both algorithms with various parameters is a good way to determine the efficiency.
The efficiency of the resource usage means the ratio of the idle state to the active 
state of processors during the optimization .

\singlespacing

\chapter{JavaSpaces Technology}

\section{JavaSpaces and Distributed Application}

\doublespacing

Software applications will change very suddenly and 
noticeably as devices become ubiquitous, network-connected, 
and ready to communicate. As the applications changes, 
the way in which one designs and builds software will 
change as well: Distributed applications involving multiple
processors and devices will become the natural way to build systems 
\cite{Freeman1999}.

However, designing distributed software is difficult.
The main characteristics of a networked environment 
(such as heterogeneity, partial failure, and latency) 
and the difficulty of {\it gluing together} multiple, 
independent processes into a robust, scalable application present the 
programmer with many challenges that do not occur when designing and 
building desktop applications.

JavaSpaces technology is a simple and powerful tool 
that makes creating distributed applications easy.
Processes are loosely coupled, communicating and synchronizing their 
activities using a persistent object store called an object space, rather
than using direct communication. It can be used to store the system state
and implement distributed algorithms. This method of coordinating
distributed applications also supports 
heavy-duty parallel computations. Space-based programming 
relies on the Jini technology's leasing\footnote{Jini\texttrademark technology is 
a service oriented architecture that defines a programming model which both 
exploits and extends Java\texttrademark technology to enable the construction 
of secure, distributed systems consisting of federations of well-behaved network 
services and clients. For this master thesis, Jini Technology Starter 
Kit v2.1 is used.}, distributed event, and transaction features, making it 
suitable for building robust, high-quality distributed 
systems \cite{Bishop2002}.

\section{Object Space}

Object Space is a new model for developing distributed  
applications. All processes of the distributed application
share an Object Space. An Object Space is a logical entity.
A service provider expresses the service as an object:
write into, read and withdraw from the space. Clients
request the object through the required service.

In JavaSpaces, all objects must also implement 
the Entry interface in which all objects are 
derived from the base class Object.
Entries can be complex or a very simple that represent unique identities in
Object Spaces. 

Object Spaces, as a computing paradigm, was put forward by 
David Gelernter at Yale University. Gelernter developed a 
language called Linda to support the concept of global 
object coordination \cite{Freeman1999}.

Object Space can be thought of as a virtual repository, 
shared amongst service providers and clients of a network, 
which are abstracted as objects. Processes 
communicate among each other using these shared objects.

An object located in a space needs to be 
registered with an Object Directory in the Object Space. 
Any processes can identify the object from the 
Object Directory, using properties lookup, where the 
property specifies the criteria for the lookup of 
the object. A process may choose to 
wait for an object to be placed in the Object Space, 
if the required object is not available.

Objects located in a space are passive, whereas the methods 
inside the objects cannot be invoked while the objects are 
in the Object Space. Instead, the processes who needs to invoke
methods in the requested object must retrieve it from the 
Object Space into its local memory, use the
service provided by the object, update the state by invoking
its method or public fields of the object 
and put it back into the Object Space \cite{Mamoud2005}.

Object Space ensures mutual exclusion. If an object is accessed, 
it has to be withdrawn from the Object Space, and will be placed back 
after it has been finished. Therefore, no other processes can access an object 
while it is being used by one process. 

\begin{figure}[htp]
\centering
\includegraphics[scale=1.0]{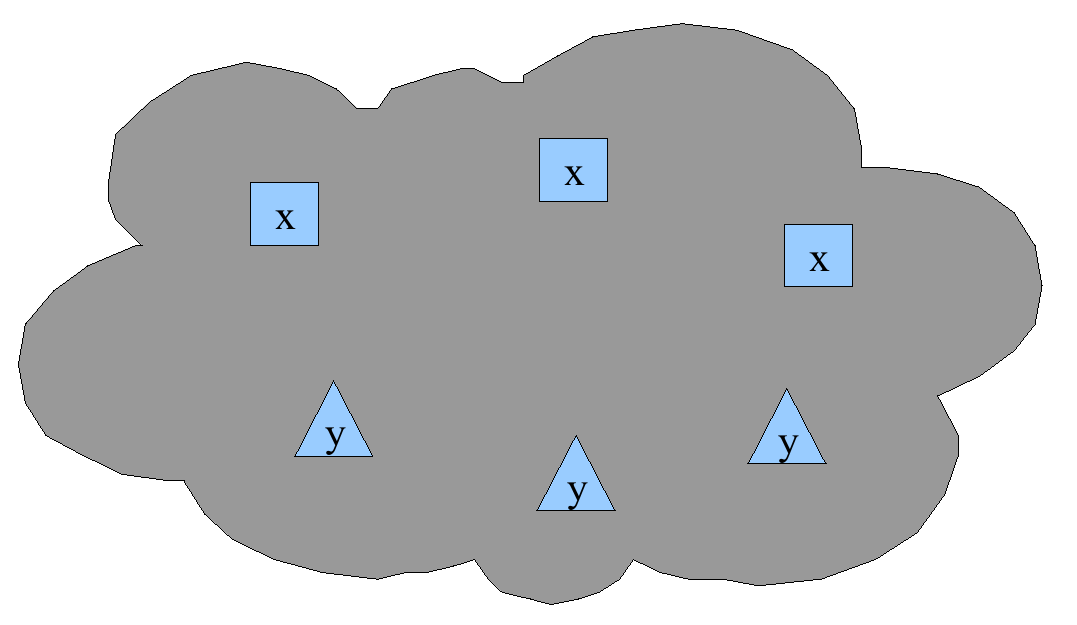}
\caption{An Object Space contains six objects with two different types: {\it rectangle} and {\it triangle}}
\label{ObjectSpace01}
\end{figure}

\begin{figure}[htp]
\centering
\includegraphics[scale=1.0]{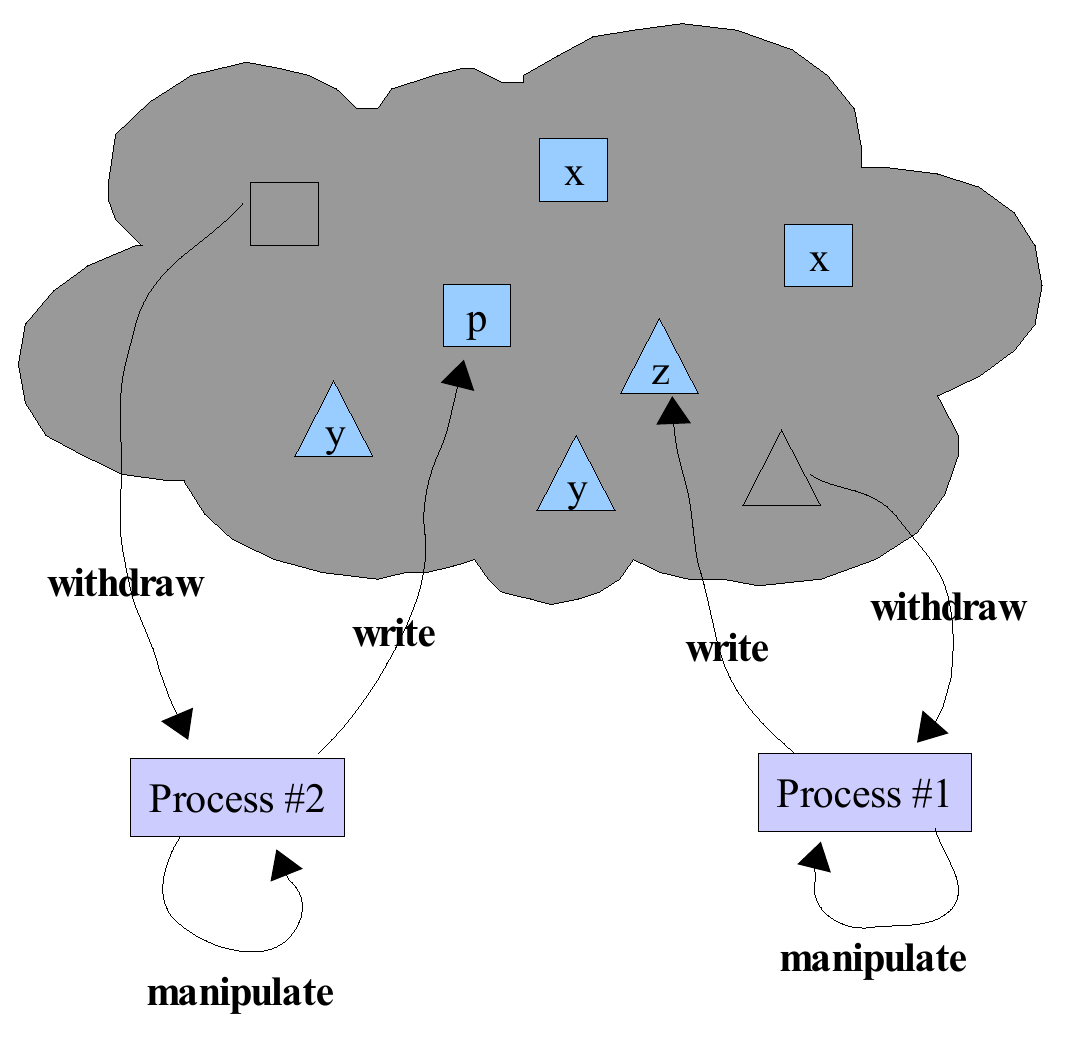}
\caption{Two processes manipulate the objects concurrently}
\label{ObjectSpace02}
\end{figure}

As shown in Figure \ref{ObjectSpace01}, an Object Space
contains six objects with two different types which are {\it rectangle} and
{\it triangle}. Consider that each {\it rectangle} object contains a value $x$ and
{\it triangle} object contains $y$. In Figure \ref{ObjectSpace02}, two processes
manipulate the objects. Assuming that both processes are running concurrently,
process one (depicted as Process \#1) takes a {\it triangle} object and alters its 
value to $z$, while process two takes a {\it rectangle} object and alters its 
value to $p$. In more details, \cite{Freeman1999,Bishop2002,Mamoud2005} 
should help the understanding how to get started with JavaSpaces Technology.

\singlespacing

\chapter{Application Design} \label{chapter3}

\doublespacing

\section{SEPP}
SEPP is a software developed by Manuel Prinz for our department 
\cite {Prinz2006}. SEPP is an abbreviation of {\it System for 
Evolutionary Pareto Optimization of Peptides/Polymers} and used to optimize 
peptide structures. SEPP is written in the Java Programming language. Moreover,
several external programs also involve in SEPP, which are:

\begin{itemize}
\item
APBS (Advanced Poisson-Boltzmann Solver) \cite{apbs},
\item
Gromacs (GROningen MAchine for Chemical Simulations) \cite{gromacs}, being
a toolbox for many programs,
\item
PDB2PQR \cite{pdb,pdb2pqr},
\item
WhatIF \cite{whatif},
\end{itemize}

\noindent
and some other packages such as BioJava to manipulate peptide structures 
\cite{biojava}.
                                     
\section{Master/Worker Technique}
\begin{figure}[htp]
\centering
\includegraphics[scale=1]{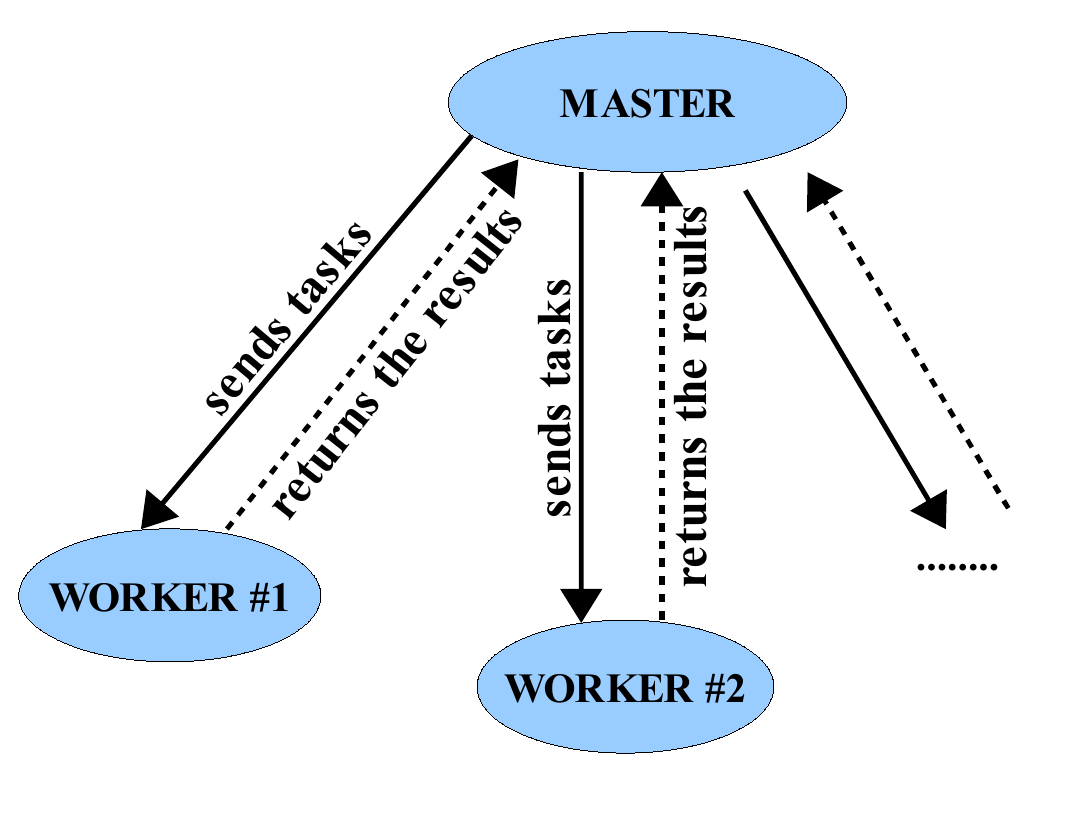}
\caption{Scheme of Master/Worker Technique}
\label{MasterWorkerScheme}
\end{figure}

SEPP is an software developed using Parallel Programming. It can run from start 
to finish on multiple resources, which correspond to processors. In SEPP, 
the processing is broken up into parts.  
The instructions from each part can run concurrently on 
different processors. The resources can exist on a single machine, or they can 
be processors in a set of computers connected via a network as shown on 
Figure \ref{MasterWorkerScheme}.

Consider that one task takes $t$ time units of execution time on a single 
processor. So if this task is broken up into $n$ parts and each of which can be 
executed on $n$ processors concurrently, the execution time then 
becomes $\frac{t}{n}$ time units. The Master/Worker Technique is used to perform 
this parallel computation \cite{Dean2004}. 
The common implementation can be seen as follows,

\noindent
MASTER:
\begin{itemize}
\item
initializes tasks and splits it up according to the number of available WORKER
\item
sends its split task to each WORKER
\item
receives the results from each WORKER
\end{itemize}

\noindent
WORKER:
\begin{itemize}
\item
receives the split task from the MASTER
\item
performs computation on the given task
\item
returns the results to the MASTER
\end{itemize}

\noindent
The Master/Worker Technique implements static load balancing which is usually 
used if all tasks perform the same amount of work on identical machines. An 
overview of how the Master works can be seen in the following pseudo-code:

\lstset{language=java}
\lstset{commentstyle=\textit}
\singlespacing
\begin{lstlisting}[frame=trbl]{}
/* MASTER */
DOP = n; 

SplitTask = Task / n;

for(i = 1; i <= n; i++) {
  send SplitTask to each WORKER;
}

while(not all tasks finished) {
  wait and receive the results from each WORKER;
}
\end{lstlisting}
\doublespacing

\noindent
The degree of parallelism (DOP) is a metric which indicates how many processes 
are being executed simultaneously. Because one processor is 
responsible for one process, the tasks should be split with respect to the 
number of processors.

\noindent
And how the WORKER works can be seen also in the following pseudo-code:

\lstset{language=java}
\lstset{commentstyle=\textit}
\singlespacing
\begin{lstlisting}[frame=trbl]{}
/* WORKER */

while(TRUE) {
  wait and receive tasks from MASTER;
  compute tasks;
  return the result to MASTER;
}
\end{lstlisting}
\doublespacing

\section{Timed Petri Nets}

\begin{figure}[htp]
\centering
\includegraphics[scale=1]{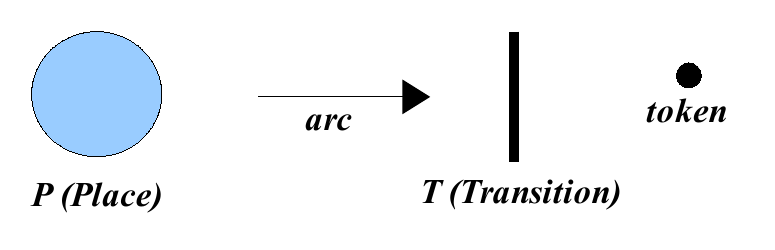}
\caption{Petri net components}
\label{PetriNet_Basic}
\end{figure}

Petri nets were invented in 1962 by Carl Adam Petri. A Petri net is one of 
several mathematical representations of discrete distributed systems. As a 
model Petri net depicts the structure of a distributed system in a graphical 
form as shown in Figure \ref{PetriNet_Basic}. It consists of {\it place}, 
{\it transition}, and directed {\it arc}.
Place is represented by circle and transition by bar. Place may contain
zero or more {\it tokens}, drawn as dot, and change during the execution of
the net. Consider that $P$ is an {\it input place} of a transition $T$ if there
exists a directed {\it arc} from $P$ to $T$; $P$ is an {\it output place} 
of $T$ if there exists a directed {\it arc} from $T$ to $P$.

\begin{figure}[htp]
\centering
\includegraphics[scale=1]{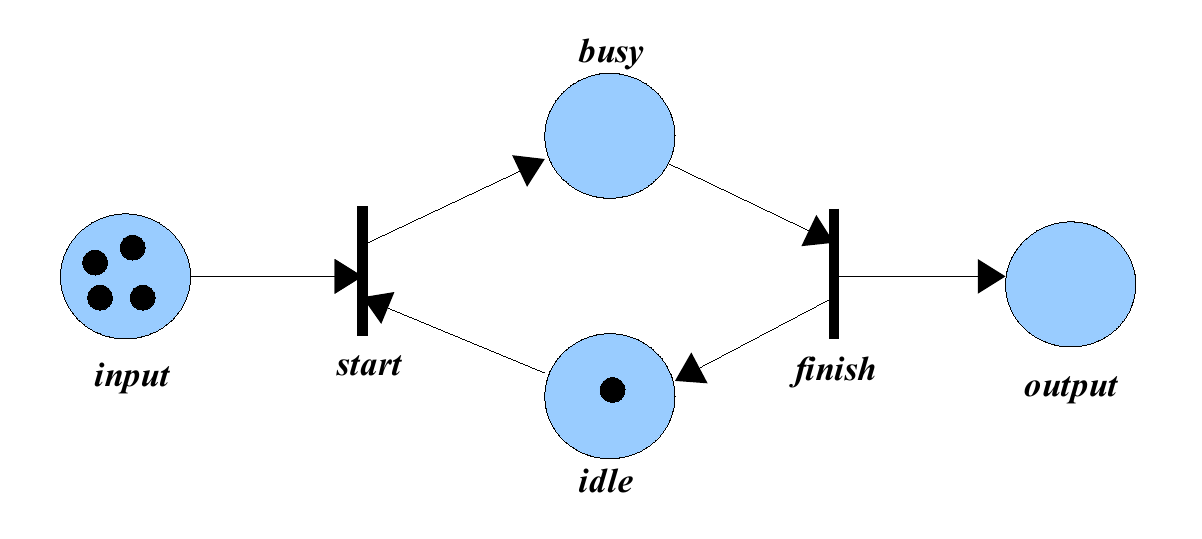}
\caption{One resource represented by a Petri net}
\label{PetriNet_01}
\end{figure}

Figure \ref{PetriNet_01} shows a classical Petri net model to illustrate the 
states of a resource \cite{Aalst}. This Petri net models a resource which 
executes or processes tasks and has two states: {\it idle} and {\it busy}. 
The state shown in Figure \ref{PetriNet_01} expresses that the resource is idle
or free. There are four tokens in {\it input place} which represent 
jobs to be executed by one resource (processor). The token in place {\it idle}
indicates that the resource is free and able to process a task.

The tokens in {\it input place} will be placed into {\it output place} if 
all tasks had been processed by the resource. Transition {\it start} has
two input places ({\it input} and {\it idle}), while transition {\it finish}
has only one input place ({\it busy}) and two output places ({\it output} 
and {\it idle}). A transition is called {\it enabled} if the condition is 
fulfilled that each of its input places contains at least one token. An 
enabled transition can {\it fire} which means transition $T$ consumes tokens 
from its input places and producing tokens for its output places.

\begin{figure}[htp]
\centering
\includegraphics[scale=1]{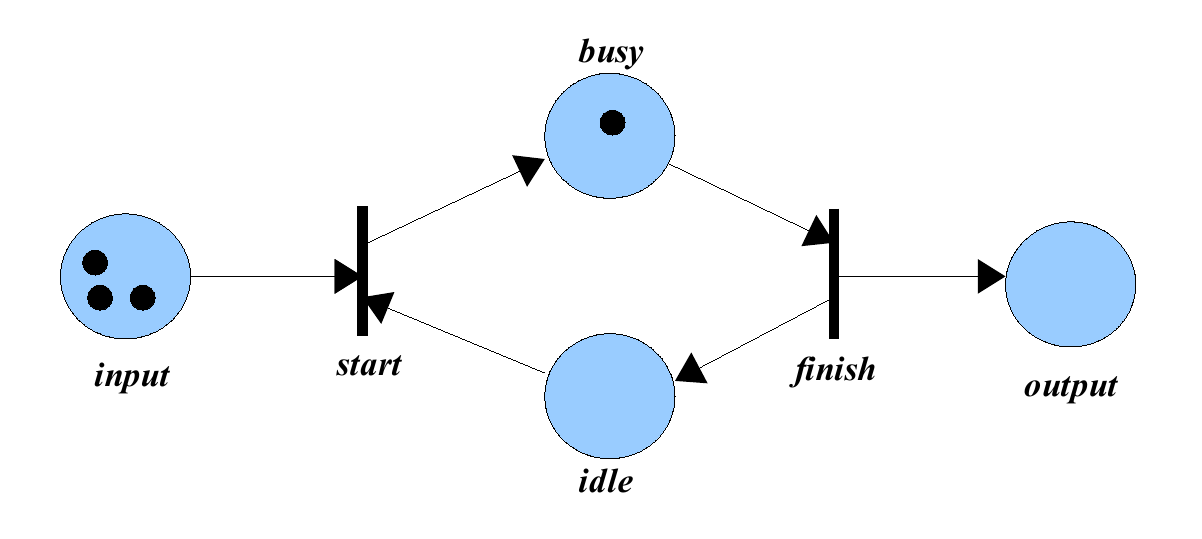}
\caption{Firing transition {\it start}}
\label{PetriNet_02}
\end{figure}

In Figure \ref{PetriNet_01} transition {\it start} is enabled to {\it fire}
because the condition is fulfilled that place {\it input} and {\it idle} 
contain at least one token. Transition {\it finish} is not enabled
because there are no tokens in place {\it busy}. Firing transition 
{\it start} means consuming two tokens, one from place {\it input} and
the other one from place {\it idle}. Hence, it produces one token for place
{\it busy}, as seen in Figure \ref{PetriNet_02}. 

\begin{figure}[htp]
\centering
\includegraphics[scale=1]{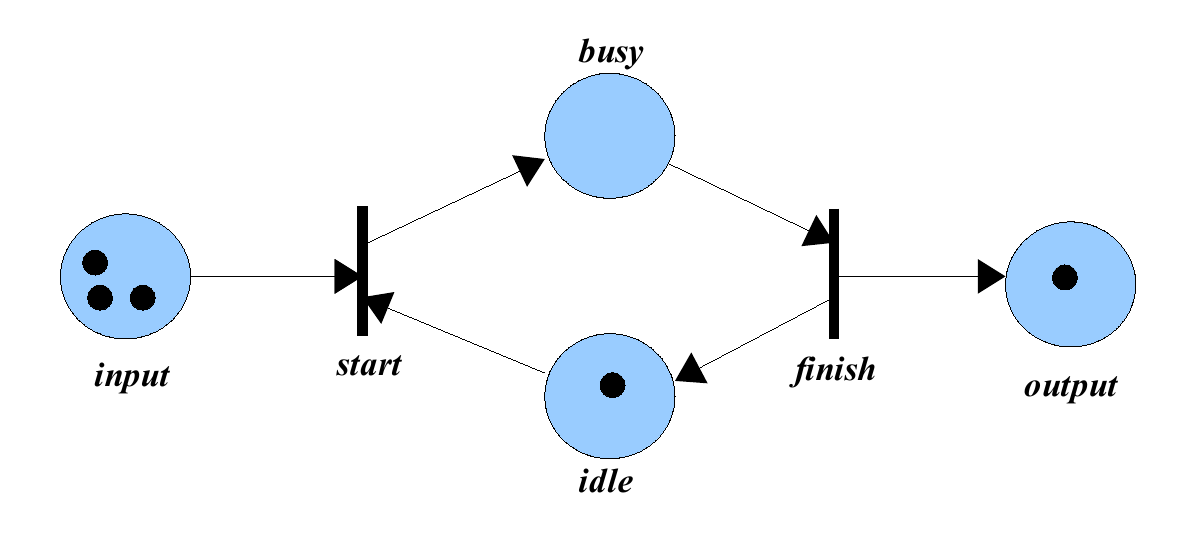}
\caption{One task has been already executed}
\label{PetriNet_03}
\end{figure}

In this state, shown in Figure \ref{PetriNet_02}, transition {\it finish} 
is enabled and transition {\it start} is disabled. Once transition 
{\it finish} has fired, the token in place {\it busy} is consumed and
two tokens are then produced: one token is for place {\it idle} and the
other one for {\it output}. Now transition {\it start} is enabled, and 
so on and so forth. As long as there are tasks waiting to be 
processed, those two transitions fire alternately because this Petri net
model can only process one task at a time. The resulting state is shown in
Figure \ref{PetriNet_03}.

\subsection{Time Attributes}

\begin{figure}[htp]
\centering
\includegraphics[scale=1]{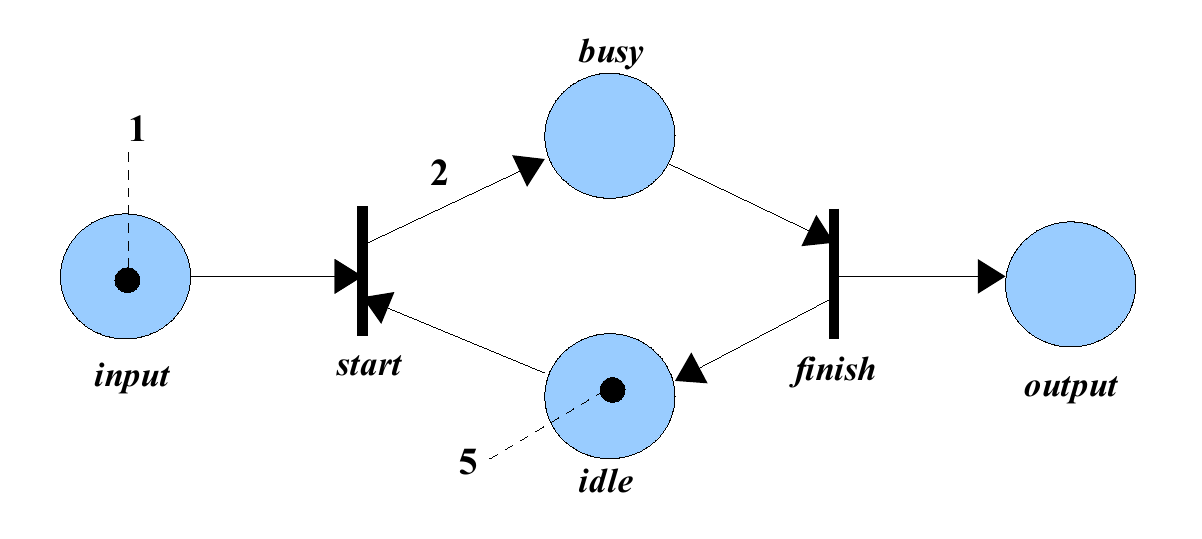}
\caption{Petri net with time attributes}
\label{PetriNet_01a}
\end{figure}

Since the classical Petri net is not easily capable of handling quantitative 
time, a timing concept was added \cite{Aalst}. Each token has 
{\it timestamp} which represents availability for consumption. Timestamps 
indicate when tokens become available and when a transition becomes enabled 
for which each of its input places contains available tokens. Moreover, 
timestamp is also equal to the {\it firing time} plus the {\it firing delay}
of the corresponding transition. Consider Figure \ref{PetriNet_01a}, place
{\it input} contains one token with timestamp 1, place {\it idle} contains 
a token with timestamp 5 and the firing delay is 2 time units. The transition 
{\it start} becomes enabled at time 1; the token being produced for place 
{\it busy} has then timestamp $1+2+5=7$.

\begin{figure}[htp]
\centering
\includegraphics[scale=0.8]{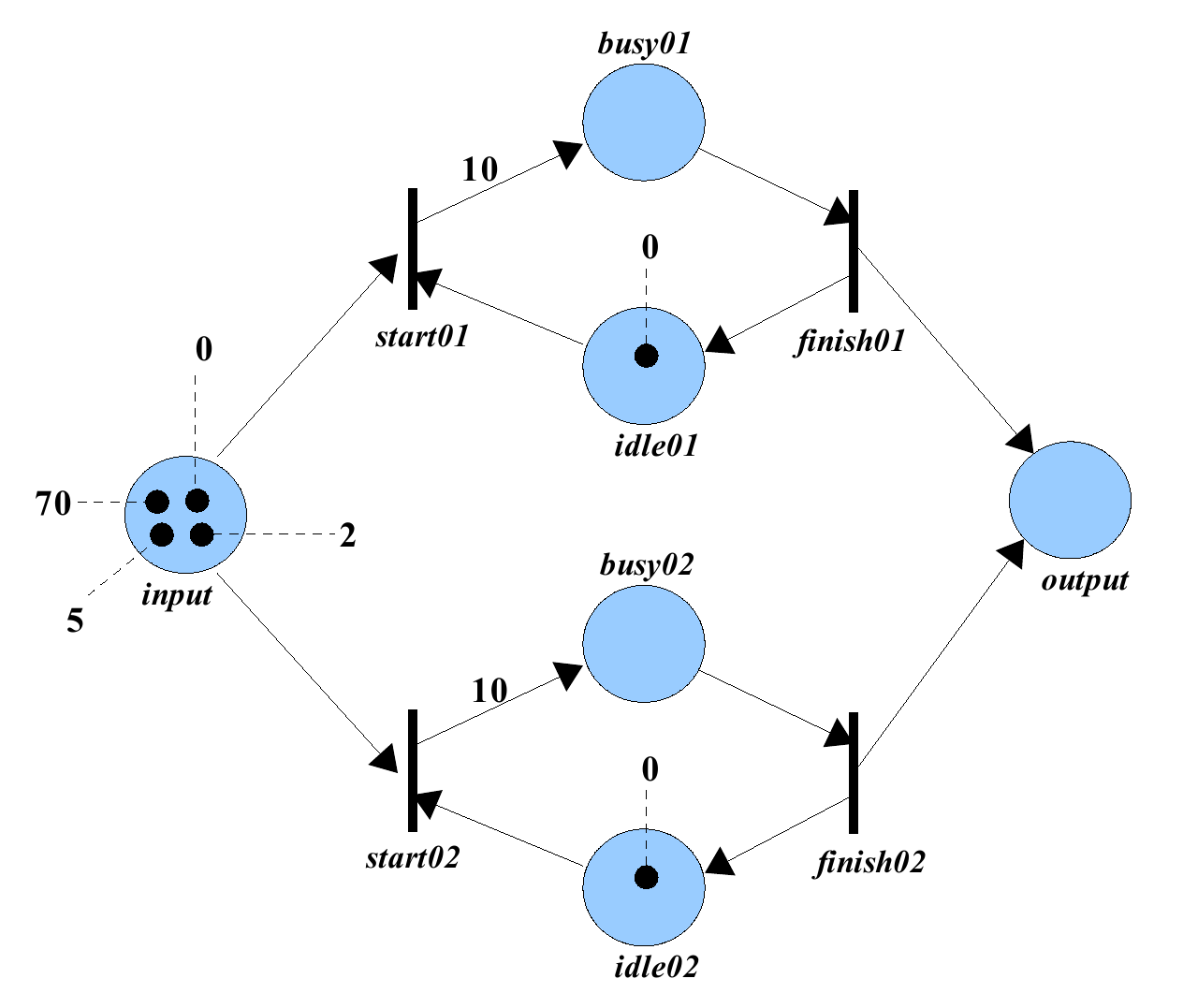}
\caption{Two identical resources}
\label{PetriNet_04}
\end{figure}

A more complex example is shown in Figure \ref{PetriNet_04}. Place {\it input} 
contains four tokens with timestamps which represent tasks. These tasks need to 
be processed by the two resources. The firing delay of transition
{\it start01} is 10, the first task arrives at time 0, the second at time 2,
the third at time 5, and the forth at time 70. The tokens in place {\it idle01}
and {\it idle01} have timestamp 0. 

\begin{figure}[htp]
\centering
\includegraphics[scale=0.8]{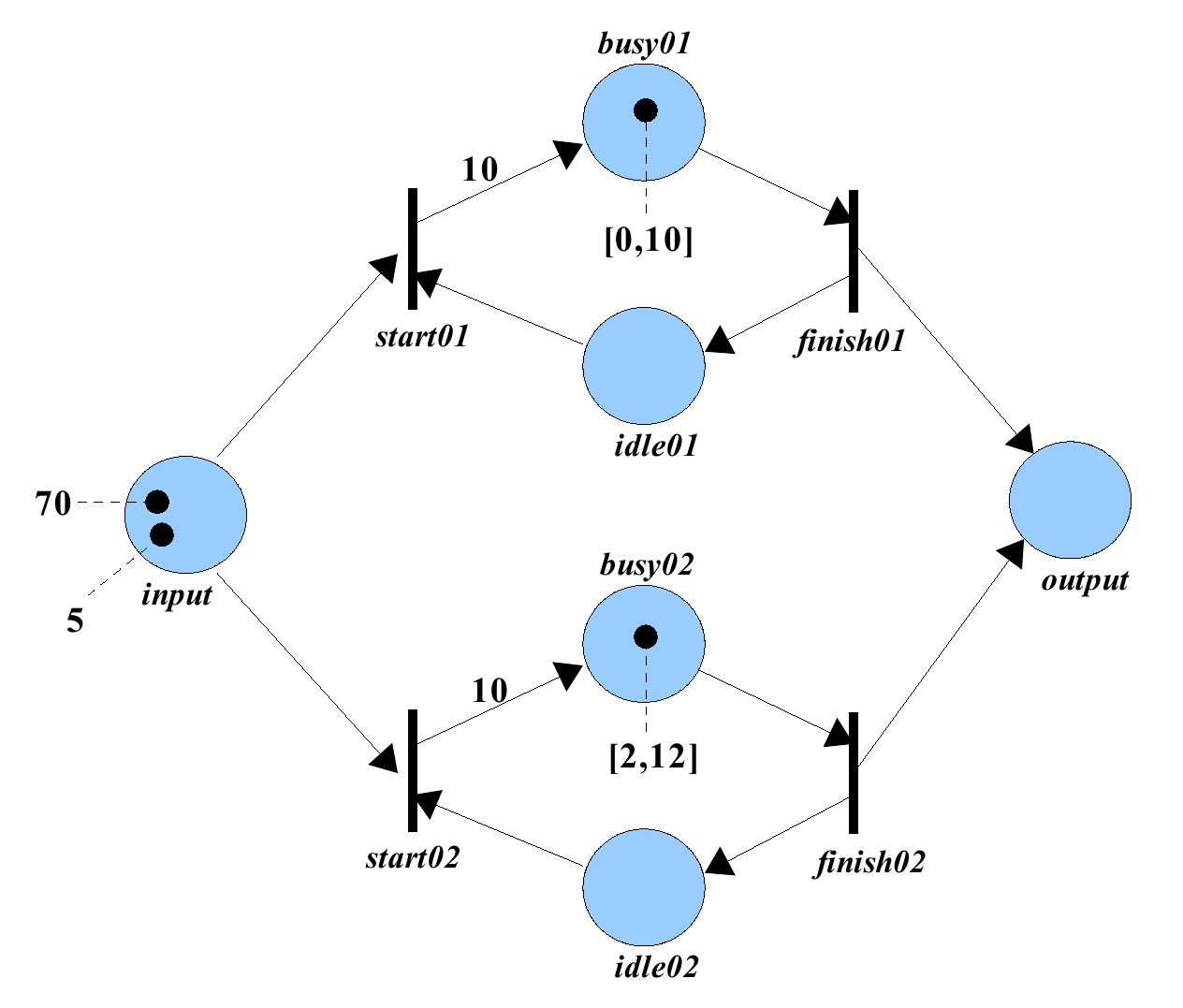}
\caption{Firing transition {\it start01} and {\it start02}}
\label{PetriNet_05}
\end{figure}

Figure \ref{PetriNet_04} tells that both resources are ready to process
a task at time 0. Therefore, one of these resources will start processing a task
that arrives at time 0. Let's assume that the first resource always takes care
of the task first. Transition {\it start01} fires at time 0 and produces a token 
with timestamp 10 time units. Notice that this firing delay represents a 
{\it processing time} of a task which is the time required to execute the task 
given a specific resource. The intermediate state from firing transition 
{\it start01} and {\it start02} is shown in Figure \ref{PetriNet_05}.

\begin{figure}[htp]
\centering
\includegraphics[scale=0.8]{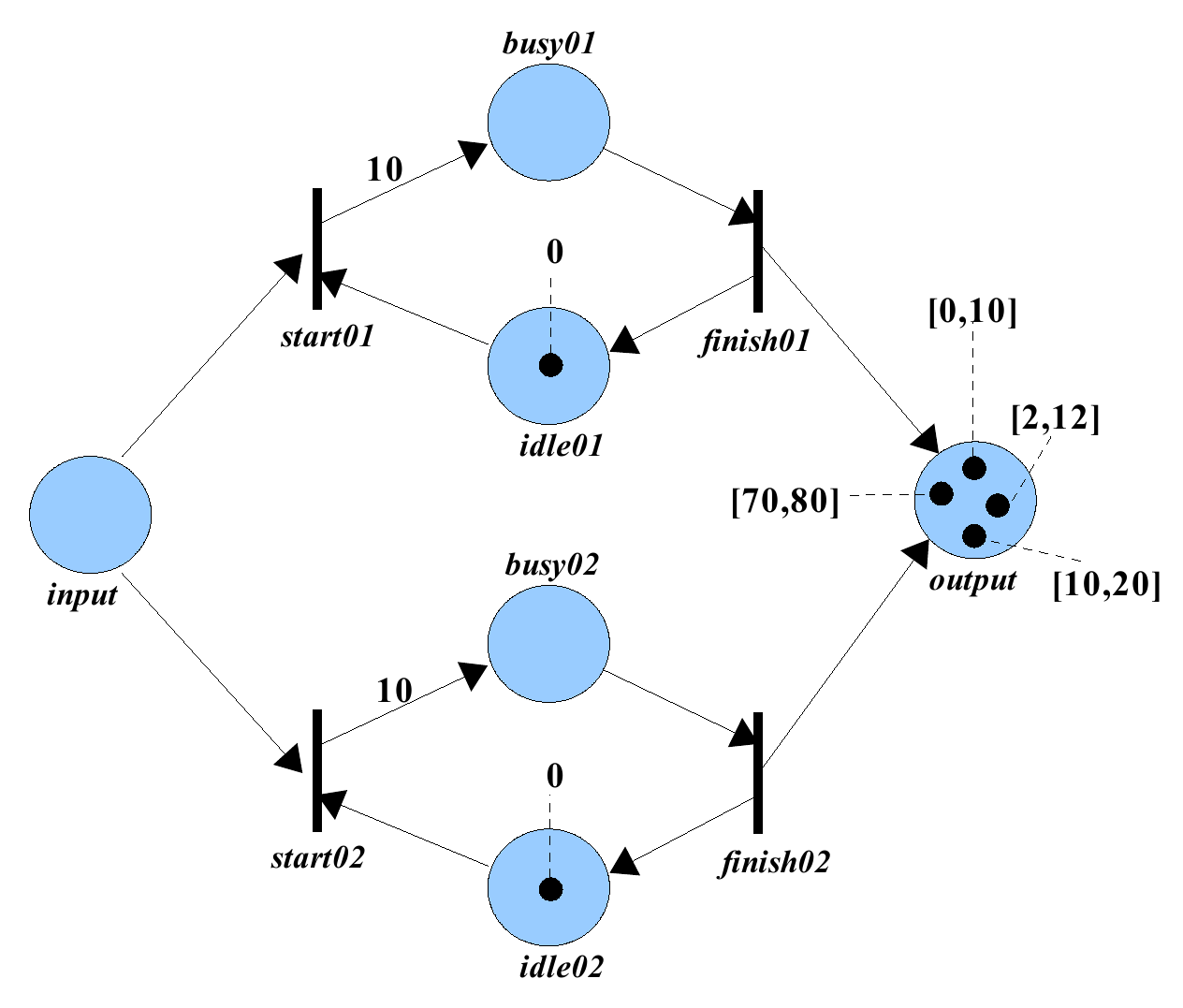}
\caption{The resulting state after all tasks have been processed}
\label{PetriNet_06}
\end{figure}

At time 2 transition {\it start02} becomes enabled. It means that the 
second resource starts to process the task arriving at time 2. Now both
resources are busy until one of them has finished processing a task. Finally, 
the first resource will take care of the task with timestamp 5 at time 10,
because from time 5 until 10 this resource was still busy, and the the task with 
timestamp 70 at time 70. Figure \ref{PetriNet_06} shows the resulting state 
from firing transition {\it start01} three times and {\it start02} once.

\begin{figure}[htp]
\centering
\includegraphics[scale=1]{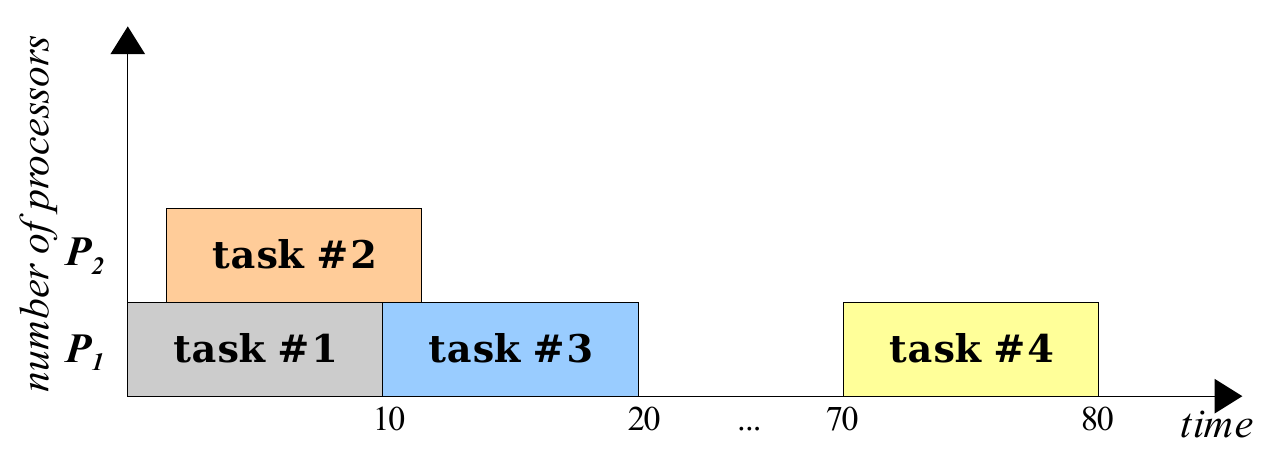}
\caption{Resource usage over time derived from the resulting state}
\label{ProcAnalysis01}
\end{figure}

Each token in place {\it output} contains a vector timestamp. The first 
element in the vector expresses the time in which one of two transitions start to
process the corresponding token; the second element apparently expresses the 
finishing time. As a result, Figure \ref{ProcAnalysis01} shows the overview of 
the resource usage over time derived from the resulting state in 
Figure \ref{PetriNet_06}. It's obvious that the resource usage is not
well-distributed; $P2$ has processed less task and has more idle state comparing
to $P1$.

\section{Timed Petri Nets and JavaSpaces Technology}

The next question is how to implement such Petri nets with time attributes
in a program. JavaSpaces Technology is one possibility to answer this question.
As described in the previous chapter, JavaSpaces Technology is a simple and 
powerful tool that makes creating distributed applications easy. 
Because Petri nets model such distributed application, then JavaSpaces 
Technology is a reasonable choice.

\begin{figure}[htp]
\centering
\includegraphics[scale=1]{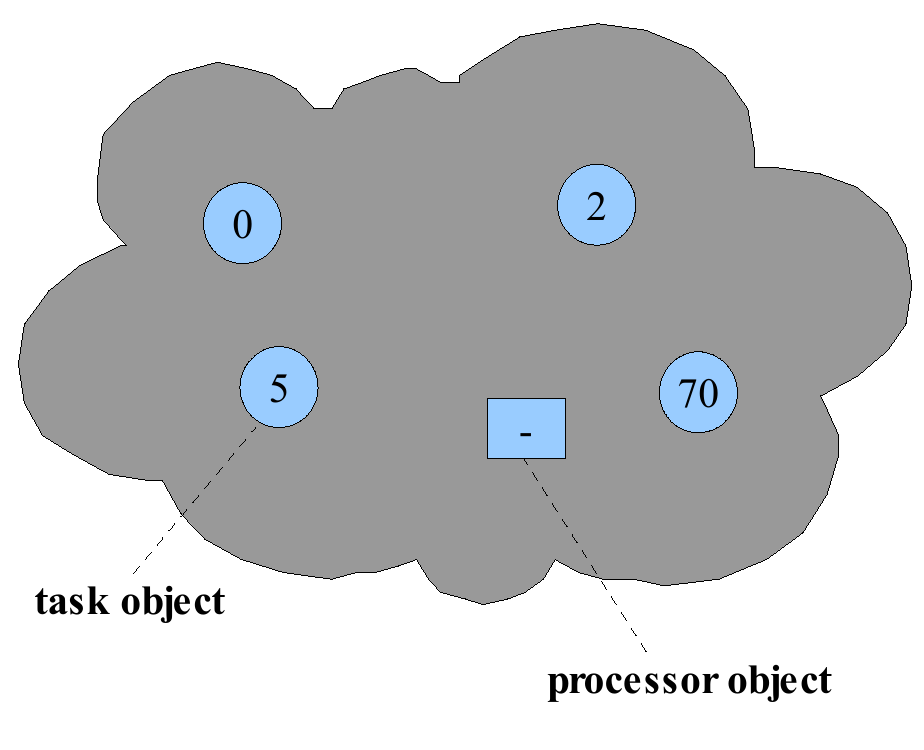}
\caption{An Object Space containing four task objects and one processor object}
\label{ObjectSpace03}
\end{figure}

In Figure \ref{ObjectSpace03}, an Object Space containing objects is depicted 
to describe the idea.
Let's consider that in the Object Space, there are four tasks with timestamps and
one resource (processor). This model is analogous to the model as shown in 
Figure \ref{PetriNet_04} with a small difference. In the Object Space, there is 
only one resource available, whereas in the previous example there are two identical
resources.
This situation models a computer with one processor and four tasks.
These tasks need to be processed using one resource. It's absolutely
certain that there will be no parallelism.

\begin{figure}[htp]
\centering
\includegraphics[scale=1]{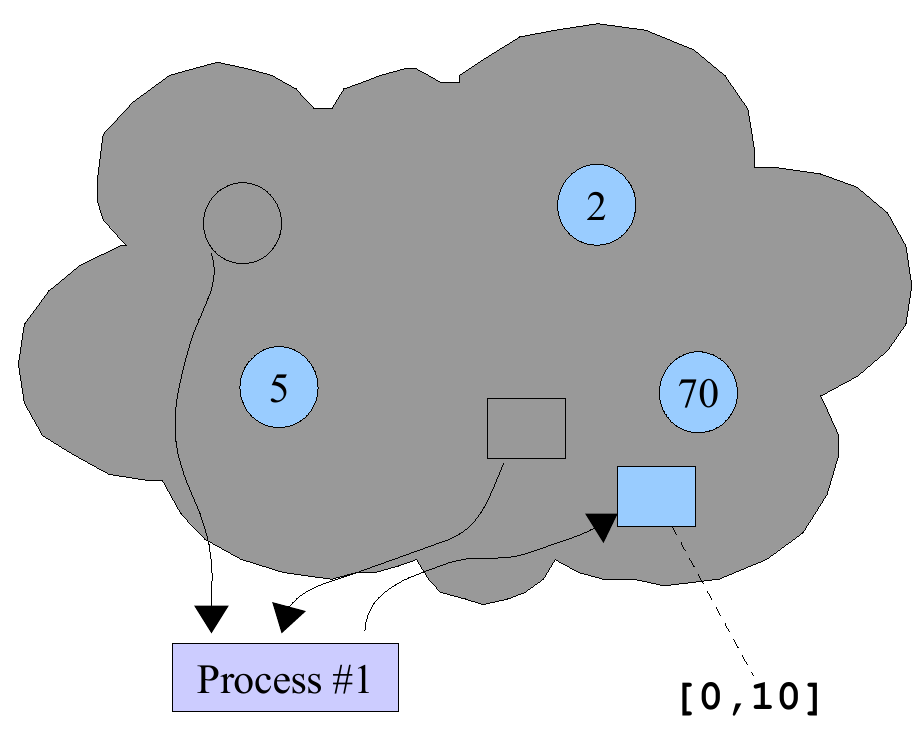}
\caption{The execution process for the first task}
\label{ObjectSpace04}
\end{figure}

In the Object Space, task object contains an unique ID and timestamp 
indicating availability for processing. As explained, a task with
timestamp $t$ means that this task arrives at time $t$ and is able
to be executed at least at time $t$. The processor object contains
an array which later contains vector timestamps expressing the starting and 
finishing time. As an additional constraint in the program, the processing time
of each task is 10 time units. 
Figure \ref{ObjectSpace04} depicts an execution process of the first task and
also shows that a timestamp $[ 0,10]$ is stored in the processor object. It means
that the execution starts at time 0 and finishes at time 10.
Notice that the second task which actually should be executed at time 2
must wait until the first task has been executed. Consequently,
the execution for the second task starts at time 10. This condition is also analogous
to the third task, nevertheless the fourth task is executed at time 70.

\begin{figure}[htp]
\centering
\includegraphics[scale=1]{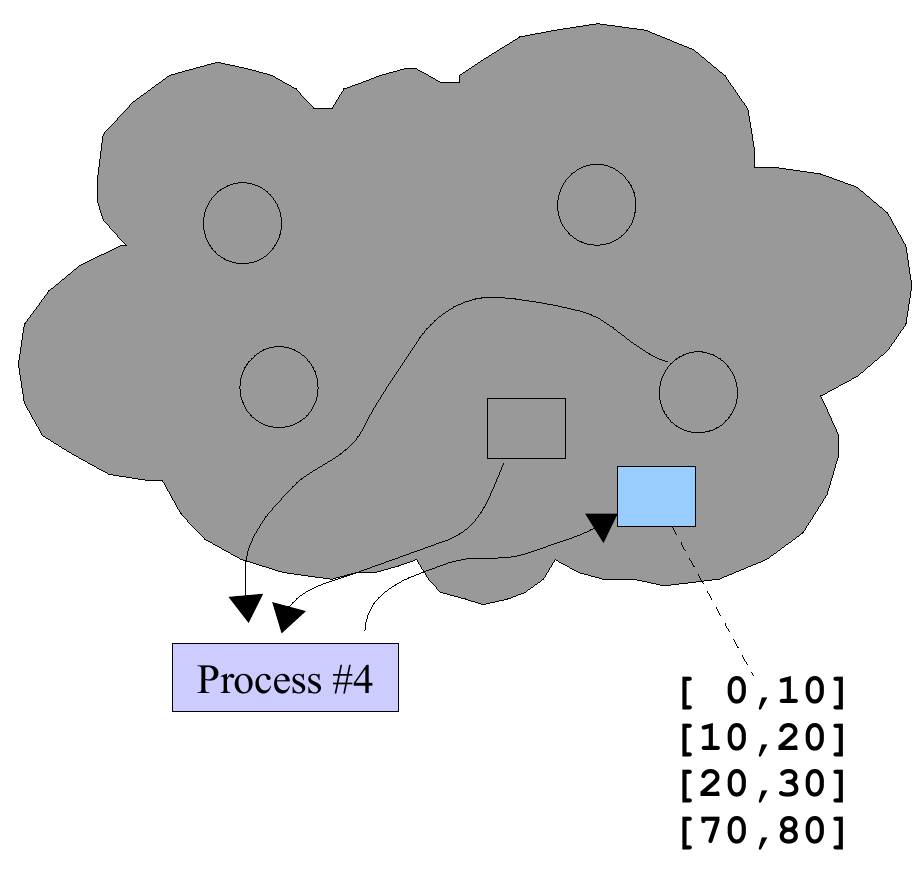}
\caption{The result after four execution processes finished}
\label{ObjectSpace05}
\end{figure}

In the program, the time unit is implemented using a real time unit. For instance, 
to process each task, it needs 10 time units which can be implemented
using real time units: {\it milliseconds}, {\it seconds}, etc.
Eventually, Figure \ref{ObjectSpace05} depicts the final situation.
It is shown that all tasks are already executed and only the processor
object is located in the Object Space. The processor object contains 
four vector timestamps in its array. As a result, Figure \ref{ObjectSpace05} 
shows the overview of the resource usage over time derived from
the result which is shown in Figure \ref{ObjectSpace04}. It's certainly
analogous for modelling a parallel computation using multiple resources.
For further use, the simulation program is developed based on this idea. 
In more details, it is explained in the next chapter on page \pageref{simulationprogram}.

\begin{figure}[htp]
\centering
\includegraphics[scale=1]{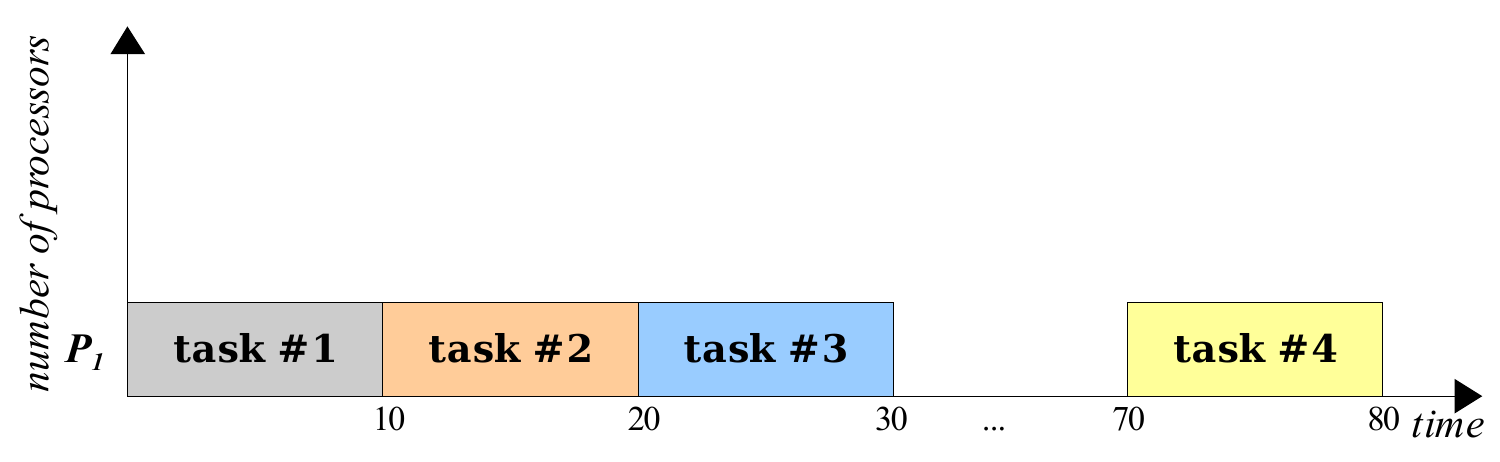}
\caption{Resource usage over time derived from the result}
\label{ProcAnalysis03}
\end{figure}

\singlespacing

\chapter{Simulation, Results, and Analysis}

\doublespacing

This chapter contains the main results on which this master thesis is
focused.

\section{Simulation Parameters}

\begin{figure}[htp]
\centering
\includegraphics[scale=0.75]{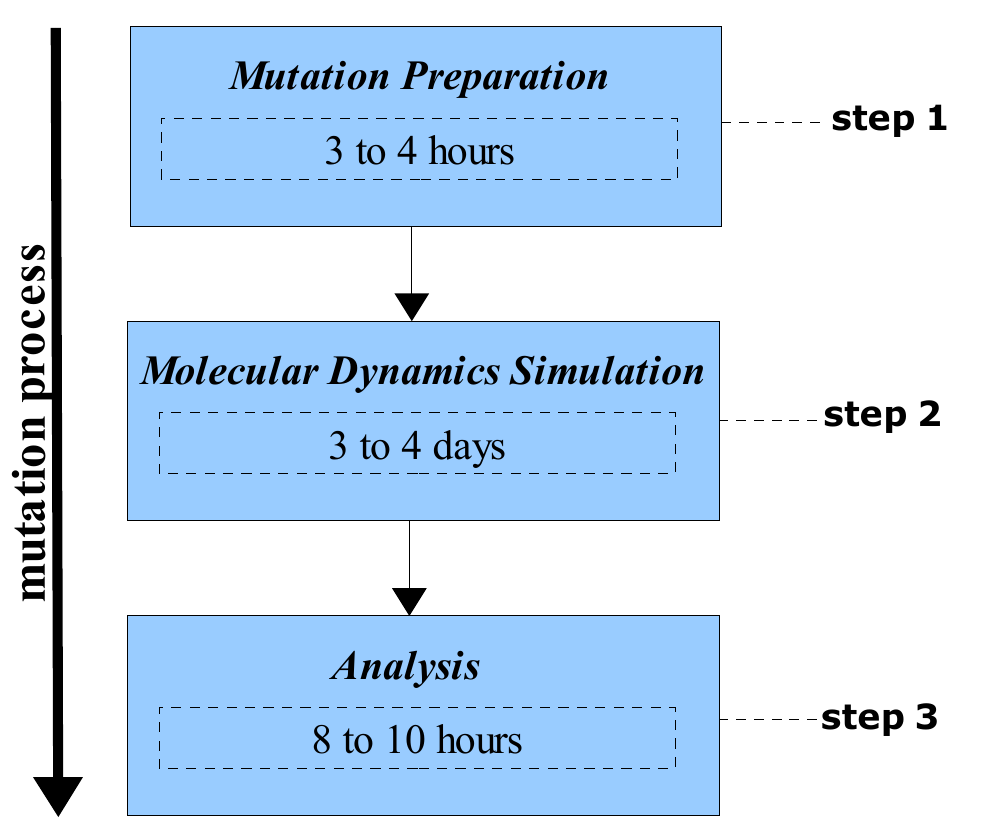}
\caption{Three computational steps need to produce a new individual
from a given individual}
\label{ComputationSteps}
\end{figure}

Figure \ref{ComputationSteps} shows the computation steps producing a new 
individual from a given individual and the whole steps are called mutation
process. The computation time for each step is taken from the real experiment 
in the computer laboratory\footnote{the computer laboratory is located at
the department I work for. This information is given by Manuel Prinz 
and Prof. Dr. Daniel Hoffmann and both are my thesis supervisors.}.
These steps are a simplification from the actual experiment with respect 
to the significance of the computation time.
It means that only computations which need to be repeated and take several
hours are considered. Hence, the other computations are hereby neglected.

In the simulation, those computation times are proportionally scaled in order
to make the simulation feasible. In this case the scale is

\begin{equation}
\centering
1~\mathrm{millisecond~(real~time)} : 1000~\mathrm{milliseconds~(simulated~time)}
\label{scalefactor}
\end{equation}

\noindent
The reason why the scale factor is $1000$ explained on page \pageref{why1000}.

Now consider that $\eta$ is the number of given individuals, $N$ is the number of
produced individuals (new individuals), $\Phi$ is the number of processors and 
$\delta_i$ is the degree of parallelism of step $i$. These notations are used 
as parameters for the simulation. Moreover, computation time for each step is 
denoted by a random variable $X$ which is normally distributed\footnote{the simulation
program had actually done a simple model based on uniformly distributed random variables,
but later changed to normal distribution because it is more realistic
according to the {\it central limit theorem}.}.
A normal distribution in a variate $X$ with mean $\mu$ and standard deviation $\sigma$ 
is a statistical distribution with the following probability function

\begin{equation}
\centering
P(x) = \frac{1}{\sigma\sqrt{2\pi}} \exp\left(-\frac{(x-\mu)^2}{2\sigma^2}\right), x \in \Re
\label{probabilityfunction}
\end{equation}

\noindent
Table \ref{timerange} contains the time range for each step with the corresponding 
notation. Notice that the time range\footnote{because negative values lead to an 
exception (no negative value for time), 
the simulation program only generates values that are natural numbers.} 
for each step is based on the real experiment, as shown in Figure \ref{ComputationSteps}, 
which is scaled according to the scale factor from Equation \ref{scalefactor}. 

\begin{table}[htp]
\centering
\begin{tabular}{c c c c c}
\hline\hline
Step & Time Range (in milliseconds) & $\mu$ (mean) & $\sigma$ (standard deviation)  & Notation\\ [0.5ex]
\hline
1 & $[10800,...,14400]$ & 12600 & 3600 & $t_{s1}$ \\
2 & $[259200,...,345600]$ & 302400 & 86400 & $t_{s2}$ \\
3 & $[28800,...,36000]$ & 32400 & 7200 & $t_{s3}$ \\ [1ex]
\hline
\end{tabular}
\caption{Time range for each computational step}
\label{timerange}
\end{table}

\section{Generation-based vs. Steady-state Algorithm}

Before running the simulation, both algorithms are statistically compared 
without considering parallelism. Let's assume that the processors are identical. 
Table \ref{table01} contains the detail of parameters. Notice that all values 
of $\delta$ are equal to 1, therefore one individual corresponds to one 
processor. Consider that the given individuals are computed\footnote{"computed" 
means "mutated".} simultaneously because $\eta$ is less than or equal 
to $\Phi$ ($\eta\leq\Phi$). 
From this it follows that the number of simultaneous mutation processes 
is equal to $\frac{N}{\eta}$. It leads to 10 simultaneous mutation processes 
that are needed to produce 100 new individuals from 10 given individuals.

\begin{table}[htp]
\centering
\begin{tabular}{c c c c c c}
\hline\hline
$\eta$ & $N$ & $\Phi$ & $\delta_1$ & $\delta_2$ & $\delta_3$ \\ [0.5ex]
\hline
10 & 100 & 10 & 1 & 1 & 1 \\ [1ex]
\hline
\end{tabular}
\caption{Parameters for statistical comparison of both algorithms}
\label{table01}
\end{table}

$T_{GB}(\eta,N)$ and $T_{ST}(\eta,N)$ are the functions describe
the overall computation time for Generation-based and Steady-state Algorithm 
respectively. The Equation \ref{T_GB} and \ref{T_ST} show the details of 
both functions. 

\begin{equation}
\centering
T_{\mathrm{GB}}(\eta,N) = \sum_{i=1}^{n} 
max\left(\vec{X_1}(i) + \vec{X_2}(i) + \vec{X_3}(i)\right)
\label{T_GB}
\end{equation}

\begin{equation}
\centering
T_{\mathrm{ST}}(\eta,N) = \sum_{i=1}^{n} 
\mu\left(\vec{X_1}(i) + \vec{X_2}(i) + \vec{X_3}(i)\right)
\label{T_ST}
\end{equation}

\noindent
where $n$ is the number of simultaneous mutation processes with respect 
to $\eta$ and $\Phi$ which is denoted by the following notion:

\begin{equation}
n = 
\begin{cases} 
\frac{N}{\eta} & \text{if  $\eta\leq\Phi$} \\
\text{not defined} & \text{if $\eta>\Phi$}
\end{cases}
\end{equation}

\noindent
In case that $\eta$ is greater than $\Phi$ is not investigated, 
even though the simulation program is able to do that.
As an explanation for the above formulas, $\vec{X_1}(i)$, $\vec{X_2}(i)$, and
$\vec{X_2}(i)$ are vectors 
with $\eta$ elements containing random values of $t_{s1}$, $t_{s2}$, and 
$t_{s3}$ at $i$-th simultaneous mutation process respectively. 

\begin{figure}[htp]
  \begin{center}
    \subfigure[]{\label{PetriNet_07}\includegraphics[scale=0.95]{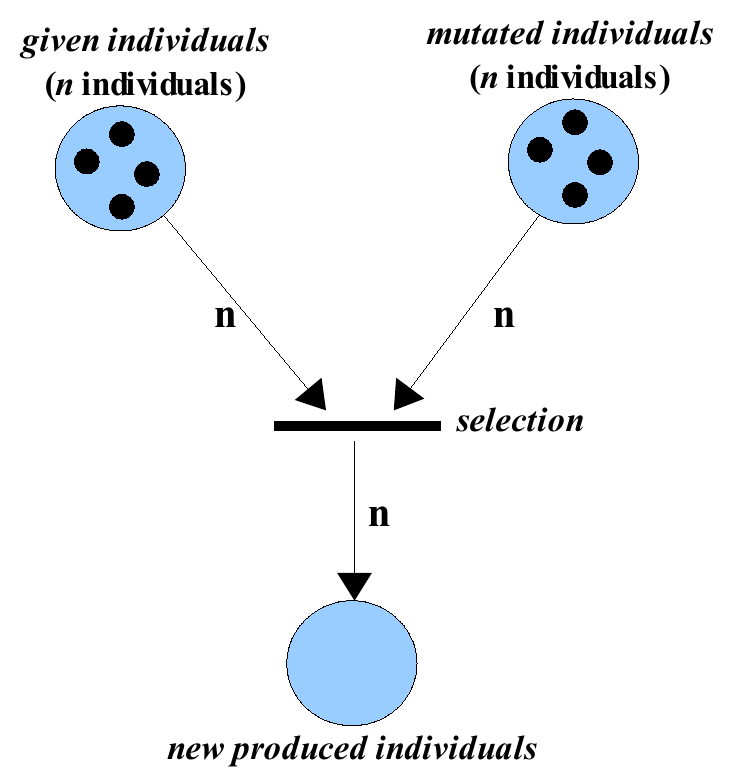}}
    \subfigure[]{\label{PetriNet_08}\includegraphics[scale=0.95]{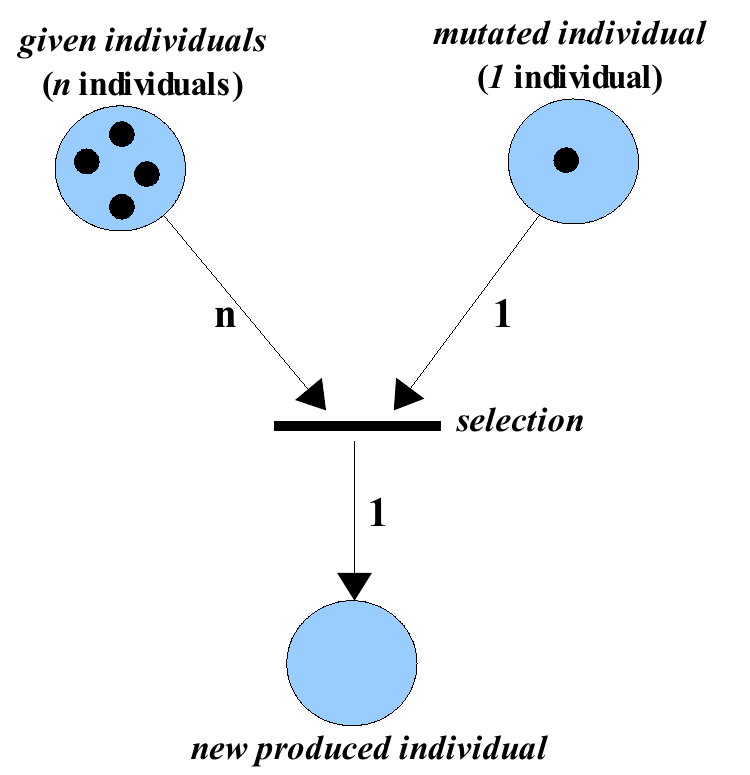}}   
  \end{center}
  \caption{Classical Petri net model shows the different type of selection 
           process for Generation-based and Steady-state Algorithm}
  \label{PetriNet_0708}
\end{figure}

As mentioned, Generation-based Algorithm 
performs selection after all mutation processes for $\eta$ individuals have been done.
Hence Equation \ref{T_GB} is reasonable to be used, in order to compute the overall 
computation time using Generation-based Algorithm.
It means that the maximum sums of $t_{s1}$, $t_{s2}$, and $t_{s3}$ 
at $i$-th to $n$-th simultaneous mutation processes are added up. 
In contrast, Steady-state Algorithm performs selections right after one individual 
has been mutated. It follows that $i$-th simultaneous mutation process is done at 
a certain time which can be 
computed by taking the mean value, denoted by $\mu$ as in Equation \ref{T_ST}, 
of the sum of $t_{s1}$, $t_{s2}$, and $t_{s3}$. Therefore $n$ mean values of 
the sum are added up in order to compute the overall computation time of 
the Steady-state Algorithm.
In Figure \ref{PetriNet_0708} the classical Petri net model shows the different 
type of selection process for Generation-based and Steady-state Algorithm.

\begin{figure}[htp]
\centering
\includegraphics[scale=0.6]{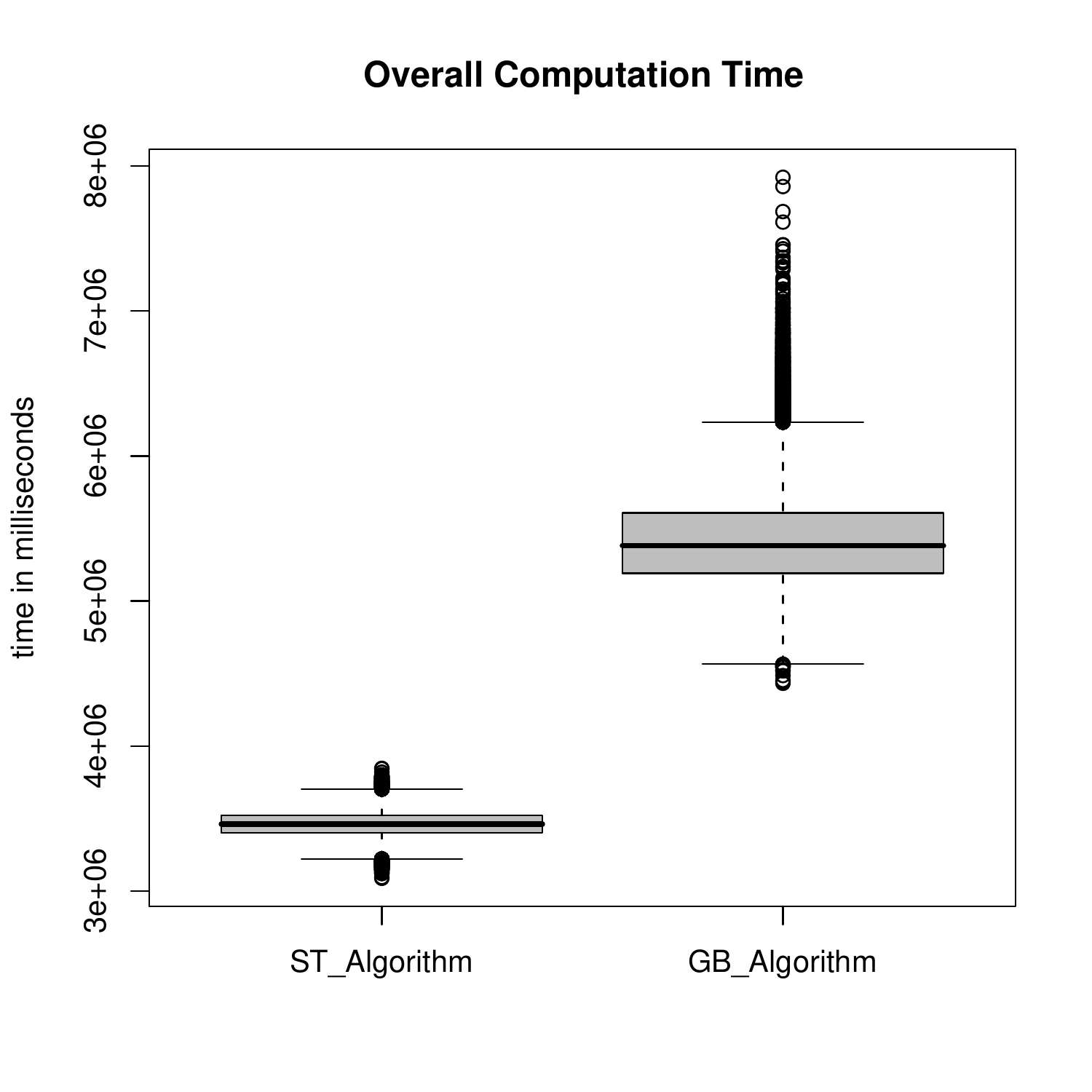}
\caption{Boxplot of statistical comparison between Generation-based and Steady-state Algorithm}
\label{boxplot_stgb}
\end{figure}

\begin{figure}[htp]
\centering
\includegraphics[scale=0.6]{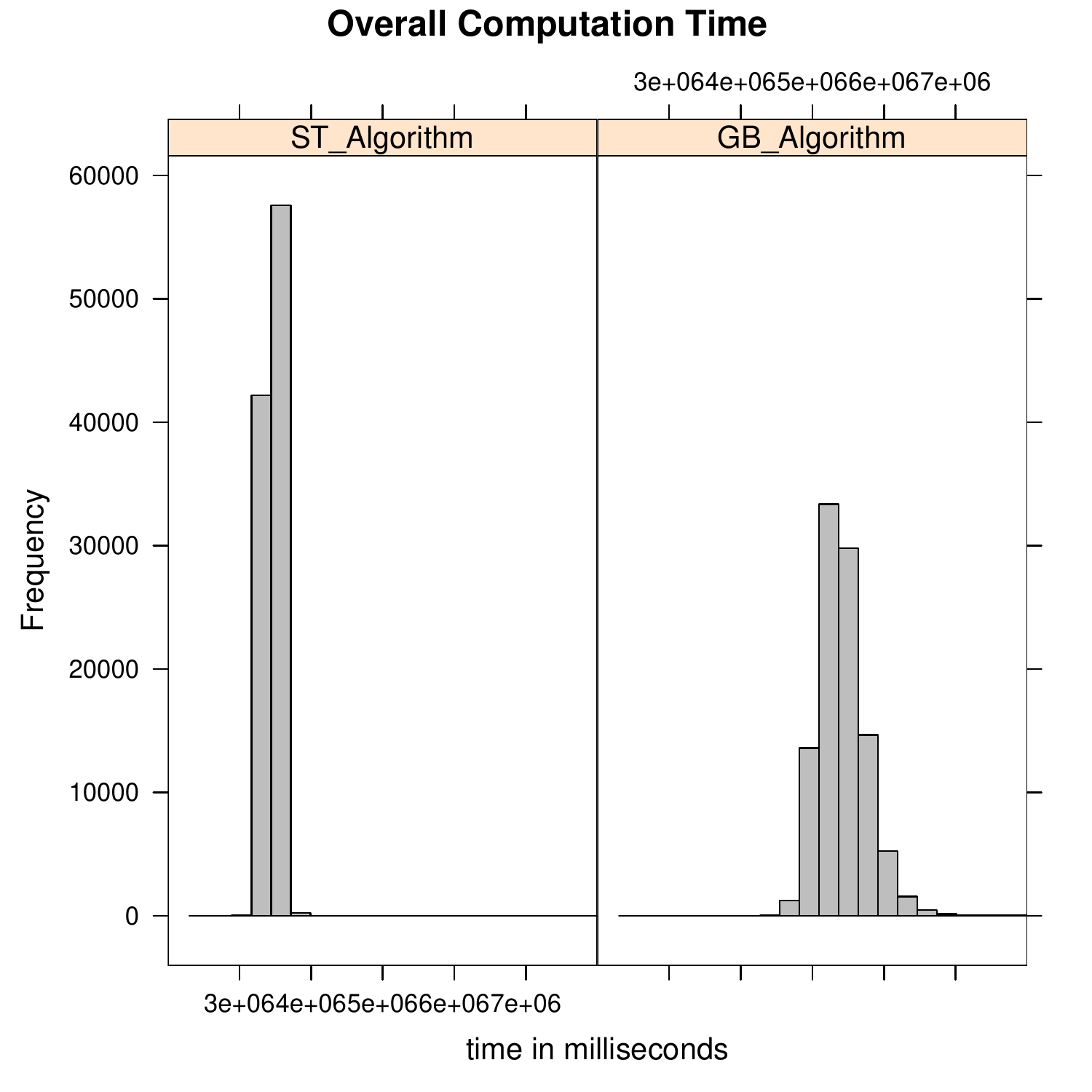}
\caption{Histogram of statistical comparison between Generation-based and Steady-state Algorithm}
\label{hist_stgb}
\end{figure}

The overall computation time for both algorithms is done based on the parameters 
from Table \ref{table01} and the Equation \ref{T_GB} and \ref{T_ST}. It is repeated
$100000$ times in order collect enough data for statistical comparison. 
Table \ref {tablestat} contains the summary of the results, whereas Figure 
\ref{boxplot_stgb} and \ref{hist_stgb} show the boxplot and histogram 
respectively.

\begin{table}[htp]
\centering
\begin{tabular}{c c c c c c c}
\hline\hline
Algorithm & Minimum & 1$^{st}$ Quantile & Median & Mean & 3$^{rd}$ Quantile & Maximum \\ [0.5ex]
\hline
ST & 3089000 & 3403000 & 3462000 & 3463000 & 3523000 & 3847000 \\
GB & 4433000 & 5190000 & 5382000 & 5417000 & 5607000 & 7921000 \\ [1ex]
\hline
\end{tabular}
\caption{Summary of the results in statistical comparison}
\label{tablestat}
\end{table}

It is now obvious that the Steady-state Algorithm is faster than the 
Generation-based Algorithm. 
To estimate {\it how much faster} that algorithm is, an approximation can be
applied as described by the following speed-up function

\begin{equation}
\centering
S(\eta,N) \approx \frac{\mu(\vec{T}_{GB}(\eta,N))}{\mu(\vec{T}_{ST}(\eta,N))}
\label{speedup}
\end{equation}

\noindent
Thus, to produce $100$ new individuals from $10$ given individuals, 
the Steady-state Algorithm is $1.56$ faster than the Generation-based 
Algorithm.

\section{Simulation Program}
\label{simulationprogram}

As explained in the previous chapter, JavaSpaces Technology is very appropriate 
tool to be used. There are three 
object types located in the Object Space. First, an individual is represented 
by an object instanced from the class 
\lstinline[basicstyle=\ttfamily,breaklines=true]!DataEntry.java!. 
Principally, this object contains an ID and the mutation index. Second, an object 
instanced from the class 
\lstinline[basicstyle=\ttfamily,breaklines=true]!ProcessorEntry.java! represents 
a processor. Processor object contains an ID and list of arrays to 
record timestamps. The last object is turn object instanced from the class 
\lstinline[basicstyle=\ttfamily,breaklines=true]!TurnEntry.java! to ensure an 
atomic collection of processors for each step. It means that the execution 
process of step $i$ starts after $\delta_i$ processors have been collected. 

The number of data (individual) and processor objects are dynamic. The values
are initialized in the class \lstinline[basicstyle=\ttfamily]!InitValue.java!.
Initialization for time range of each step is also done in this class. After
all classes compiled, the 
\lstinline[basicstyle=\ttfamily,breaklines=true]!InsertEntry.class! is run to 
place $\eta$ data objects, $\Phi$ processors and one turn object into 
the Object Space. 

This situation models a computer cluster with $\Phi$ processors and various 
tasks. The class \lstinline[basicstyle=\ttfamily]!Agent.java!
performs the computation for the individuals. This class is run using 
{\it multiple threads} which are implemented from interface 
\lstinline[basicstyle=\ttfamily,breaklines=true]!Runnable! in the class 
\lstinline[basicstyle=\ttfamily,breaklines=true]!RunAgentGenBased.java! and
\lstinline[basicstyle=\ttfamily,breaklines=true]!RunAgentSteadyState.java!. 
Those classes are run alternately depending on which algorithm is being 
simulated. According to the algorithms, maximum number of {\it threads} in
the simulation of Generation-based Algorithm is always $\eta$. Contrary to that,
more than $\eta$ {\it threads} may exist in the simulation of Steady-state
Algorithm.

In step $i$, each {\it thread} takes\footnote{"take" is equal to "withdraw".} one 
data item (individual), the turn and $\delta_i$ processor objects from the Object Space. 
As mentioned above, the turn object is needed to ensure the collection of processors.
Notice that there is no proper scheduling algorithm applied to maintain
the processor usage. Consequently the {\it thread} in step $i$ randomly takes 
$\delta_i$ processors which are available in the Object 
Space\footnote{according to SEPP application.}.
Once the processor objects have been collected, the {\it thread} 
places back the turn object immediately into the Object Space and holds the data 
and processor objects for $\frac{t_{si}}{\delta_i}$ milliseconds. Before placing 
back the objects into the Object Space, the timestamps generated by method 
\lstinline[basicstyle=\ttfamily,breaklines=true]!System.currentTimeMillis()!
are recorded in the processor objects. Moreover, the mutation index in the data
object is also incremented. 

\section{Analysis of Results}

Table \ref{simulationscenarios} contains the scenario details for the simulation. 
Scenario e00 describes the real situation.
For all scenarios there are $8$ given individuals and $80$ new produced 
individuals. Each scenario has different 
parameters for $\Phi$ and $\delta_2$. Parameter $\Phi$ or the number of 
processors plays an important role. The efficiency of
the resource usage is evaluated using various number of processors. 
As mentioned, 76 processors are based on the real situation. 
Increasing $\Phi$ up to 128 is corresponding to an extension of the cluster.

Moreover, parameter 
$\delta_2$ or the degree of parallelism of step 2 is also another important
issue\footnote{As a matter of fact, parameters $\delta_1$ and $\delta_3$ are also 
important.
But for the sake of simplicity, the simulation only considers the most dominating 
computation time which is $\delta_2$.}. 
It is used to examine the outcome when
the degree of parallelism has been increased. In order to collect enough data, 
the simulation is repeated $100$ times per scenario. It's now clear
why those scenarios, as shown in Table \ref{simulationscenarios}, are chosen. 
But notice that the simulation program actually is able to handle other variations 
as long as ${\delta_1,\delta_2,\delta_3} \leq \Phi$, because it doesn't make 
sense to have a degree of parallelism which is higher than the available processors.

\begin{table}[htp]
\centering
\begin{tabular}{c c c c c c c}
\hline\hline
Scenario & $\eta$ & $N$ & $\Phi$ & $\delta_1$ & $\delta_2$ & $\delta_3$ \\ [0.5ex]
\hline
e00 & 8 & 80 & 76 & 10 & 2 & 2 \\
e02 & 8 & 80 & 76 & 10 & 10 & 2 \\
e12 & 8 & 80 & 128 & 10 & 2 & 2 \\
e14 & 8 & 80 & 128 & 10 & 10 & 2 \\ [1ex]
\hline
\end{tabular}
\caption{Scenarios for Simulation}
\label{simulationscenarios}
\end{table}

For each simulation, the program generates the following log file:

\singlespacing
\begin{scriptsize}
 \begin{verbatim} 
  ProcID: 1; start_t: 1190380329418; stop_t: 1190380330568; data_id: 2; mutation: 1; step: 1
  ProcID: 1; start_t: 1190380330598; stop_t: 1190380331793; data_id: 8; mutation: 1; step: 1
  ProcID: 2; start_t: 1190380329418; stop_t: 1190380330568; data_id: 2; mutation: 1; step: 1
  ProcID: 2; start_t: 1190380330598; stop_t: 1190380331793; data_id: 8; mutation: 1; step: 1
  ProcID: 3; start_t: 1190380329418; stop_t: 1190380330568; data_id: 2; mutation: 1; step: 1
  ProcID: 3; start_t: 1190380330598; stop_t: 1190380331793; data_id: 8; mutation: 1; step: 1
  ProcID: 4; start_t: 1190380329418; stop_t: 1190380330568; data_id: 2; mutation: 1; step: 1
  ProcID: 4; start_t: 1190380330598; stop_t: 1190380331793; data_id: 8; mutation: 1; step: 1
  ProcID: 5; start_t: 1190380517645; stop_t: 1190380518851; data_id: 5; mutation: 2; step: 1
  ProcID: 5; start_t: 1190380704474; stop_t: 1190380705754; data_id: 2; mutation: 3; step: 1
  .....
 \end{verbatim} 
\end{scriptsize}
\doublespacing

\noindent
Notice that this log file is truncated. In fact, each log file consists of 
thousands of lines. By taking the lowest timestamp and subtracting all timestamps 
with it and some modifications, the above log file can be converted as follows:

\singlespacing
\begin{scriptsize}
 \begin{verbatim} 
  ProcID: 1; start_t:      0; stop_t:   1150; used: 1150ms; data_id: 2; mutation: 1; step: 1
  ProcID: 1; start_t:   1180; stop_t:   2375; used: 1195ms; data_id: 8; mutation: 1; step: 1
  ProcID: 2; start_t:      0; stop_t:   1150; used: 1150ms; data_id: 2; mutation: 1; step: 1
  ProcID: 2; start_t:   1180; stop_t:   2375; used: 1195ms; data_id: 8; mutation: 1; step: 1
  ProcID: 3; start_t:      0; stop_t:   1150; used: 1150ms; data_id: 2; mutation: 1; step: 1
  ProcID: 3; start_t:   1180; stop_t:   2375; used: 1195ms; data_id: 8; mutation: 1; step: 1
  ProcID: 4; start_t:      0; stop_t:   1150; used: 1150ms; data_id: 2; mutation: 1; step: 1
  ProcID: 4; start_t:   1180; stop_t:   2375; used: 1195ms; data_id: 8; mutation: 1; step: 1
  ProcID: 5; start_t: 188277; stop_t: 189433; used: 1156ms; data_id: 5; mutation: 2; step: 1
  ProcID: 5; start_t: 375056; stop_t: 376336; used: 1280ms; data_id: 2; mutation: 3; step: 1
  .....
 \end{verbatim} 
\end{scriptsize}
\doublespacing

\begin{figure}[htp]
\centering
\includegraphics[scale=1]{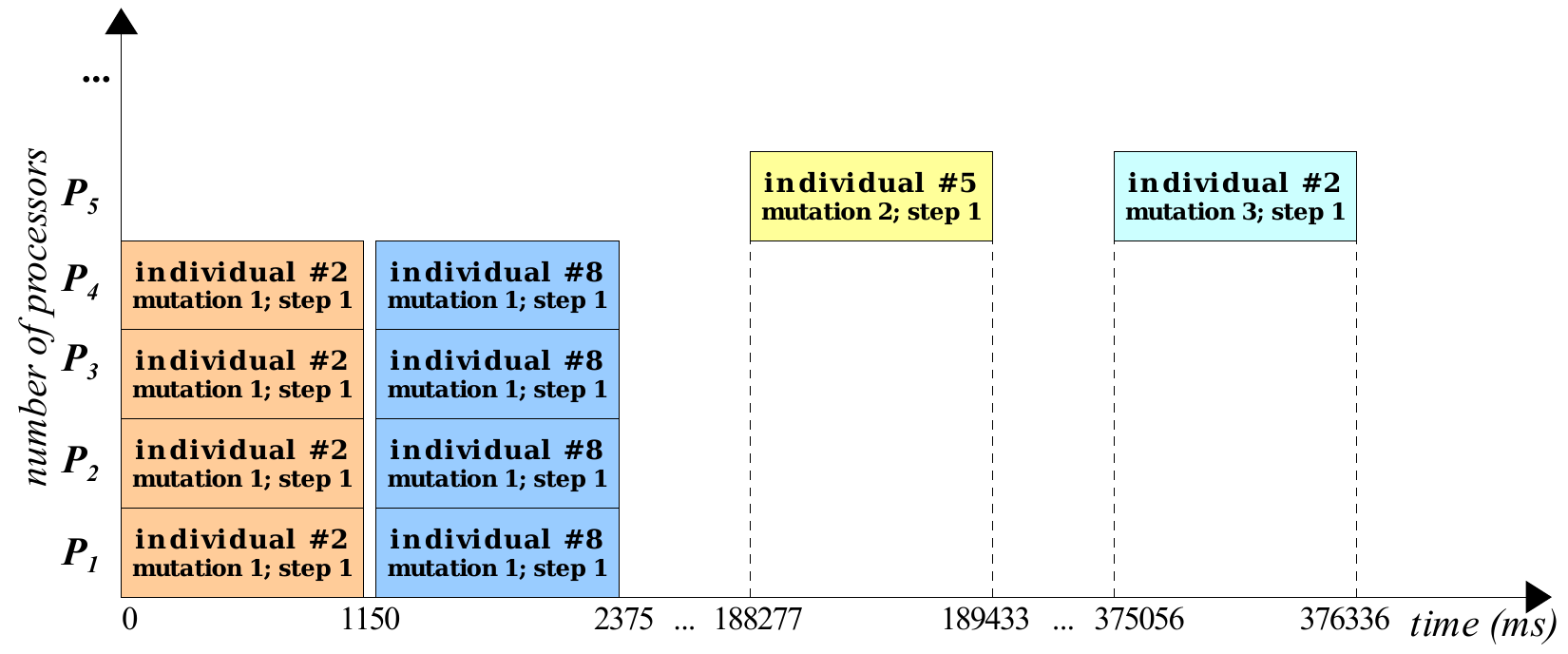}
\caption{Resource usage over time derived from the log file}
\label{ProcAnalysis02}
\end{figure}

\noindent
As a result, Figure \ref{ProcAnalysis02} shows the overview of the resource 
usage over time derived from the converted log file. It's now obvious that the
resource usage\footnote{usage of each processor.} can be determined.
In addition, the overall computation time for all scenarios can be approximated 
by subtracting the highest
timestamp with the lowest one. Consider that the highest timestamp is $\theta_h$
and the lowest is $\theta_l$, the following function, denoted by $\tau(\eta,N)$,
describes the approximation of the overall computation time.

\begin{equation}
\centering
\tau(\eta,N) \approx (\theta_h - \theta_l) - t_\mathrm{overhead} 
\label{Toverall}
\end{equation}

\noindent
It necessarily follows that the speed-up function also becomes different,
described as follows

\begin{equation}
\centering
S_s(\eta,N) \approx \frac{\mu(\vec{\tau}_{GB}(\eta,N))}{\mu(\vec{\tau}_{ST}(\eta,N))}
\label{simulationspeedup}
\end{equation}

\noindent
As an explanation, $\vec{\tau}_{GB}(\eta,N)$ and $\vec{\tau}_{ST}(\eta,N)$ are 
vectors of the overall
computation time with $100$ elements and contain the results of the
simulations based on Generation-based and Steady-state Algorithm respectively.
Table \ref{tableresults} contains the summary of the results. 

\begin{table}[htp]
\centering
\begin{tabular}{c c c c c c c}
\hline\hline
Scenario & Minimum & 1$^{st}$ Quantile & Median & Mean & 3$^{rd}$ Quantile & Maximum \\ [0.5ex]
\hline
e00$_{ST}$ & 1670000 & 1839000 & 1891000 & 1893000 & 1952000 & 2121000 \\
e00$_{GB}$ & 2102000 & 2255000 & 2314000 & 2314000 & 2364000 & 2522000 \\
e02$_{ST}$ &  488000 &  520700 &  531700 &  533100 &  544700 &  579100 \\
e02$_{GB}$ &  635400 &  675600 &  695000 &  694800 &  714200 &  770100 \\
e12$_{ST}$ & 1687000 & 1808000 & 1872000 & 1869000 & 1932000 & 2051000 \\
e12$_{GB}$ & 2101000 & 2240000 & 2287000 & 2298000 & 2351000 & 2468000 \\
e14$_{ST}$ &  490800 &  510300 &  519800 &  521700 &  528400 &  583900 \\
e14$_{GB}$ &  576000 &  603300 &  609700 &  614600 &  625400 &  691800 \\ [1ex]
\hline
\end{tabular}
\caption{Summary of the results for all simulations}
\label{tableresults}
\end{table}

\begin{figure}[htp]
\centering
\includegraphics[scale=0.5]{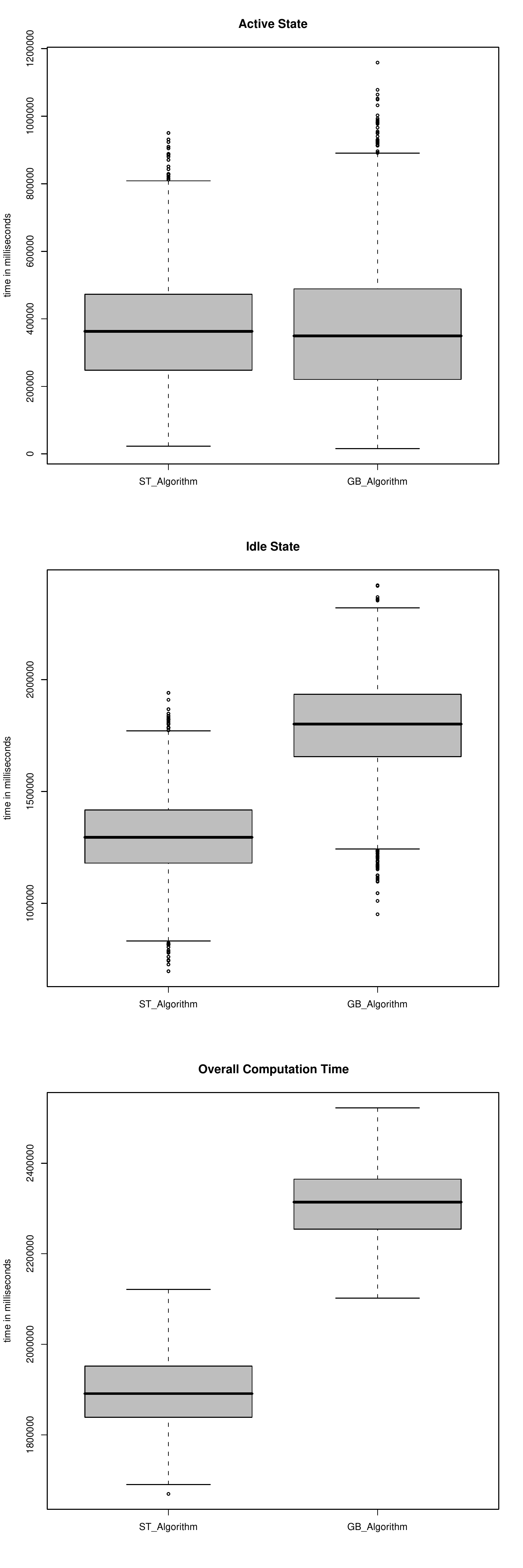}
\caption{Boxplots of scenario e00}
\label{e00}
\end{figure}

\begin{figure}[htp]
\centering
\includegraphics[scale=0.5]{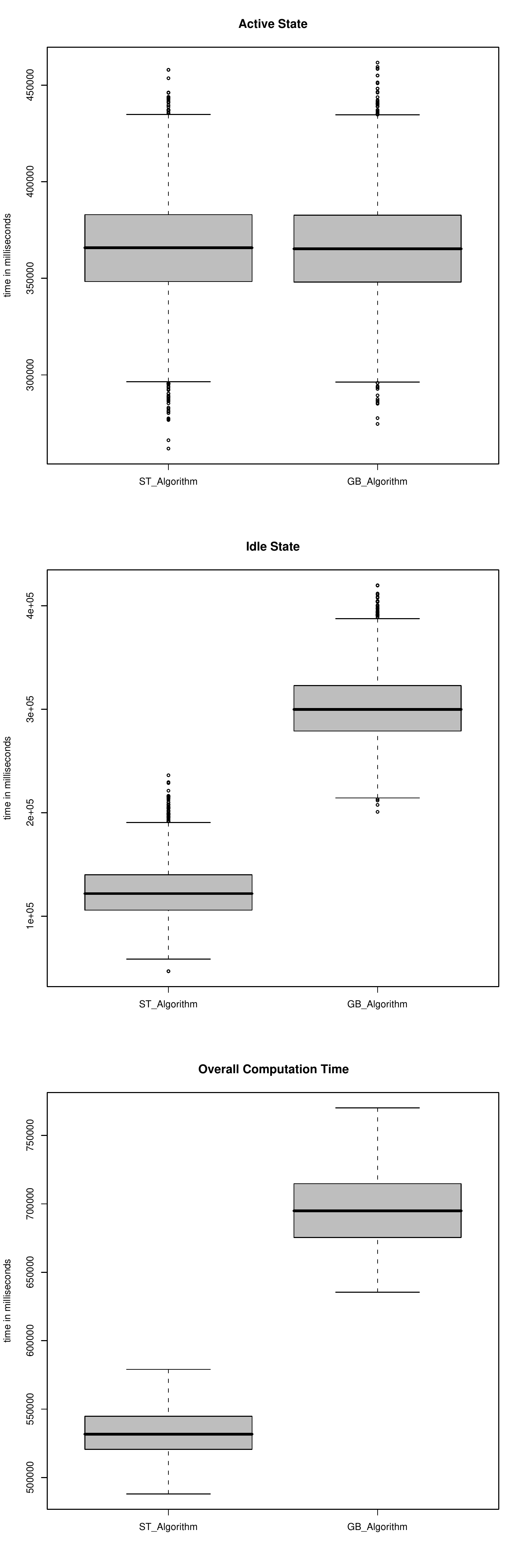}
\caption{Boxplots of scenario e02}
\label{e02}
\end{figure}

\begin{figure}[htp]
\centering
\includegraphics[scale=0.5]{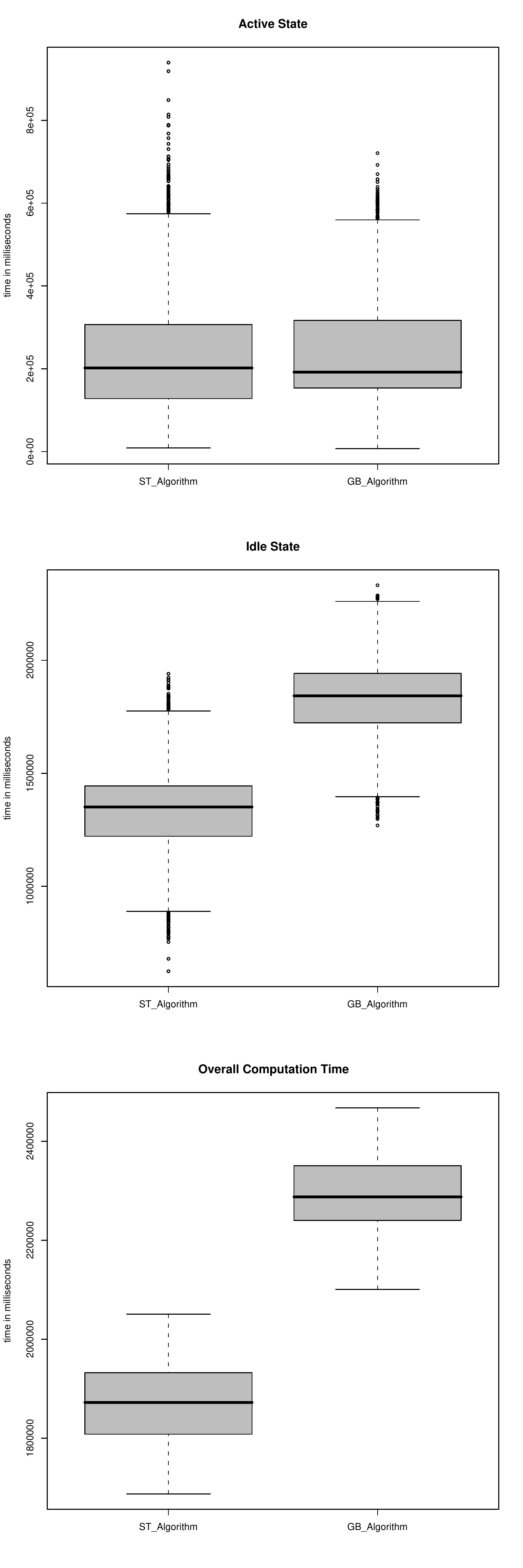}
\caption{Boxplots of scenario e12}
\label{e12}
\end{figure}

\begin{figure}[htp]
\centering
\includegraphics[scale=0.5]{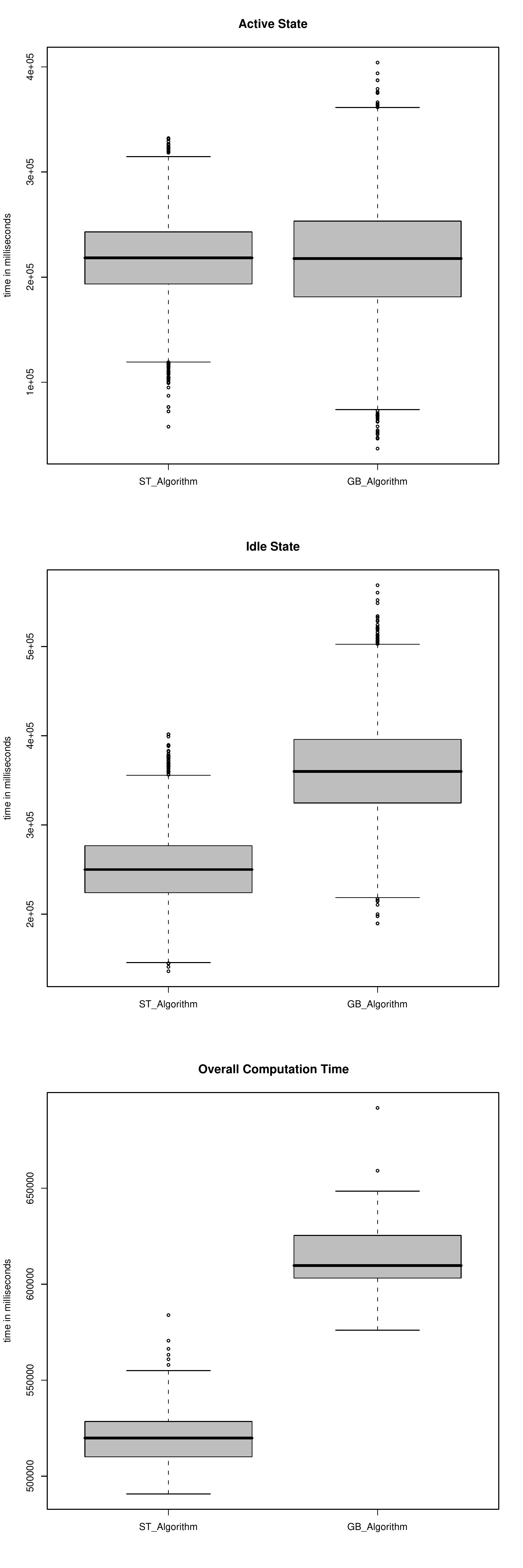}
\caption{Boxplots of scenario e14}
\label{e14}
\end{figure}

By taking into account that there exists computational overhead and
the timestamp is in milliseconds which are error-prone during the execution,
the scale factor is reduced to compensate this problem. As seen
in Equation \ref{scalefactor}\label{why1000}, the scale factor is $1000$ even 
though the higher scale factor is also possible. Consequently the ratio of 
$\frac{t_{\mathrm{overhead}}}{(\theta_h - \theta_l)}$ becomes very small and
therefore $t_{\mathrm{overhead}}$ is neglected. Furthermore,
the results, as contained in Table \ref{tableresults}, considerably make 
sense because there also exists computational overhead in the real computation. 
To be more clear, the results which are in milliseconds are converted into
days. Table \ref{overallspeedup} contains the conversion and the approximated 
speed-up. In addition, Figure \ref{e00}, \ref{e02}, \ref{e12}, and
\ref{e14} show the boxplots of all scenarios, whereas the histograms
of the overall computation time are given in Appendix \ref{appendixB}.

As an interesting thing, as seen in Table \ref{tableresults}, the number of
processor ($\Phi$) doesn't matter to the overall computation time based on
the results of scenario e00 and e12. But notice that this is only for $\eta$ 
much smaller than $\Phi$. In contrast, the degree of parallelism
($\delta_2$) plays an important role to accelerate the overall computation time.
It's also interesting to see the results of scenario e02 and e14 whose 
$\delta_2$ is five times bigger than the previous scenarios. For 
the Generation-based Algorithm, increasing $\Phi$ produces the slightly 
different results. Nevertheless, the Steady-state Algorithm yields
the better results at the overall computation time.

\begin{table}[htp]
\centering
\begin{tabular}{c c c c c c c c c c}
\hline\hline
Scenario & $\eta$ & $N$ & $\Phi$ & $\delta_1$ & $\delta_2$ & $\delta_3$ & $\mu(\vec{\tau}_{ST}(\eta,N))$ & $\mu(\vec{\tau}_{GB}(\eta,N))$ & $S_s(\eta,N)$ \\
& & & & & & & (in days) & (in days) & \\ [0.5ex]
\hline
e00 & 8 &  80 &  76 & 10 &  2 & 2 & 21.91 & 26.78 & 1.22 \\
e02 & 8 &  80 &  76 & 10 & 10 & 2 &  6.17 &  8.04 & 1.30 \\
e12 & 8 &  80 & 128 & 10 &  2 & 2 & 21.63 & 26.60 & 1.23 \\
e14 & 8 &  80 & 128 & 10 & 10 & 2 &  6.04 &  7.11 & 1.18 \\ [1ex]
\hline
\end{tabular}
\caption{Time of approximated overall computation time and speed-up}
\label{overallspeedup}
\end{table}

As a further analysis, the idle and active state of processors are also 
investigated. The processor states are inherently shown in more wide view as 
depicted in Figure \ref{e00}, \ref{e02}, \ref{e12} and \ref{e14}.
It is clearly seen that the Steady-state Algorithm is more efficient than the 
Generation-based Algorithm with respect to the resource usage.
Table \ref{usageanalysis} contains the processor states and the ratio of the 
idle to active state for all scenarios.

\begin{table}[htp]
\centering
\begin{tabular}{c c c c c}
\hline\hline
Scenario & $\mu(\vec{\tau}(\eta,N))$ & $\mu(\vec{t}_{\mathrm{active}})$ & $\mu(\vec{t}_{\mathrm{idle}})$ & $\frac{\mu(\vec{t}_{\mathrm{idle}})}{\mu(\vec{t}_{\mathrm{active}})}$ \\ [0.5ex]
\hline
e00$_{ST}$ & 1893000 & 366900 & 1298000 & 3.54 \\ 
e00$_{GB}$ & 2314000 & 364400 & 1790000 & 4.91 \\ 
e02$_{ST}$ &  533100 & 365300 &  123700 & 0.34 \\ 
e02$_{GB}$ &  694800 & 365900 &  301000 & 0.82 \\ 
e12$_{ST}$ & 1744592 & 218400 & 1526192 & 6.99 \\ 
e12$_{GB}$ & 1857025 & 218300 & 1638725 & 7.51 \\ 
e14$_{ST}$ &  521700 & 217400 &  251500 & 1.16 \\ 
e14$_{GB}$ &  614600 & 217200 &  361000 & 1.66 \\ [1ex]
\hline
\end{tabular}
\caption{Resource usage (active and idle state) for all scenarios}
\label{usageanalysis}
\end{table}

As an explanation, $\vec{t}_{\mathrm{active}}$ and $\vec{t}_{\mathrm{idle}}$ are
vectors with $\Phi$ elements containing the active and idle state of processors 
respectively. As a result of scenario e00, it is obviously shown that the 
idle state is much higher than the active state. In contrast, the active 
state is higher than the idle state for scenario e02. It explains that
the the resource usage becomes more efficient by increasing $\delta_2$.
As another fact, the efficiency of the resource usage decreases by increasing 
the number of processors from $76$ to $128$, as shown in scenario e12 and e14.

\singlespacing

\chapter{Conclusion and Future Work}

\doublespacing

\section{Conclusion}

In this chapter, some achievements from three points of view are
described as follows

\noindent
{\bf Degree of Parallelism}

The degree of parallelism plays an important role in the computation.
It accelerates the computation time and the resource usage becomes 
more efficient. Hence it concludes that a particular work to increase the 
degree of parallelism should be done.

\noindent
{\bf Number of Processors}

Increasing the number of processors is considerably not a good solution.
Many processors are not useful to speed up the computation time and 
it is expensive if the ratio of the idle 
to active state of each processor is high. It's considerably more useful
to use few processors but efficient which means that the idle
state of each processor is much smaller than the active one. 
Based on the present results, the available resources are theoretically 
possible to produce more new individuals.

\noindent 
{\bf Efficiency of the Algorithms}

As a fact of the results, the Steady-state Algorithm is better than 
the Generation-based Algorithm with respect to the computation time and 
resource usage. So it is highly recommended to use the Steady-state Algorithm
for parallel evolutionary peptide optimization.
\pagebreak

\section{Future Work}

There are possibilities to work further based on the results, such as

\noindent
{\bf Prediction about Minimum Resources}

In some facts, it's explained that increasing the number of processors 
doesn't matter to speed up the computation time. Therefore, it is 
interesting to know the minimum resources to perform the computation
producing the same results.
This prediction is useful to make the resources are efficiently used
in the real situation.

\noindent
{\bf Scheduling Algorithm}

The other possibility is implementing a scheduling algorithm. It is 
highly recommended to be implemented in order to 
maintain the resource usage and useful to increase the efficiency 
of the resource usage. If the algorithm yields the better results in
the simulation than present, it is also appropriate to be implemented
in the real situation.  In addition, it also helps to solve the problem 
if the degree of parallelism is not feasible to be increased.

\singlespacing

\begin{appendix}

\chapter{Source Code}

These are the complete source codes of my program:

\lstset{language=java, frame=tb, commentstyle=\textit, stringstyle=\upshape, showspaces=false}
\begin{scriptsize}
\begin{lstlisting}[caption=Agent.java]
/**
 * Agent.java - processing the data
 * @author: Andias Wira-Alam <andias.alam@stud.uni-due.de>
*/

import net.jini.space.JavaSpace;
import net.jini.core.lease.Lease;

public class Agent implements Runnable {
  
  private JavaSpace space;

  public Agent() {
    System.out.println("Searching for JavaSpace...");
    Lookup finder = new Lookup(JavaSpace.class);
    JavaSpace space = (JavaSpace) finder.getService();
    System.out.println("A JavaSpace has been found!");
  }

  public Agent(JavaSpace space) {
    this.space = space;
  }

  public void run() {
    try {
      
      System.out.println("Starting Agent!");
      
      ProcessorEntry ptemplate = new ProcessorEntry(); // Processor object template
      ProcessorEntry[] presult = null;
      DataEntry dtemplate = new DataEntry(); // Data object template
      DataEntry dresult = new DataEntry();
      TurnEntry turn = new TurnEntry();
      InitValue init = new InitValue();
      GenRandomTime gen = new GenRandomTime();
      long start = 0, stop = 0;
      int time_step = 0;
      boolean isDataValid = false;

      System.out.println("Taking processor and data objects from the space... ");
      
      // take 1 VALID Data from the space.
      while(!isDataValid) { 
        dresult = (DataEntry) space.take(dtemplate, null, Long.MAX_VALUE);
	if(dresult.mutationIndex.intValue() < init.numOfMutation) {
          dresult.incrementIndex(); // increment the Mutation Index.
	  isDataValid = true;
	}
	else
	  space.write(dresult, null, Lease.FOREVER);
      }
      for(int i=1; i <= init.numOfStep; i++) { // repeat acc. to the # of steps.
        presult = new ProcessorEntry[gen.getPar(i)];
        space.take(turn, null, Long.MAX_VALUE); // wait until its turn.
	for(int j=0; j < gen.getPar(i); j++) { // repeat acc. to the # of parallelism.
	  presult[j] = (ProcessorEntry) space.take(ptemplate, null, Long.MAX_VALUE);
	}
	space.write(turn, null, Lease.FOREVER); // freeing turn.
        start = System.currentTimeMillis();
	do { // avoid negative time step
	  time_step = gen.getTime(i);
	}while(time_step<0);
        Thread.sleep(time_step/gen.getPar(i)); 
        stop = System.currentTimeMillis();
        // assign values into array in the processor object.
        for(int k=0; k < gen.getPar(i); k++) { // repeat acc. to the # of parallelism
          presult[k].addRecord(start, stop, dresult.data_id.intValue(), 
	    dresult.mutationIndex.intValue(), i);
	  System.out.println("ProcID: " +presult[k].proc_id+ "; start_t: " +start+ 
	    "; stop_t: " +stop+ "; data_id: " +dresult.data_id.intValue()+ 
	    "; mutation: " +dresult.mutationIndex.intValue()+ "; step: " +i);
	  // write back the processor object.
	  space.write(presult[k], null, Lease.FOREVER);
        }
        presult = null;
      }
      // write back the Data object into the space.
      space.write(dresult, null, Lease.FOREVER);
      System.out.println("Done!");

    } catch(Exception e) {
        e.printStackTrace();
      }
  }

  public static void main(String argv[]) {
    (new Thread(new Agent())).start();
  }

}
\end{lstlisting}
\begin{lstlisting}[caption=CleanSpace.java]
/**
 * CleanSpace.java - cleaning objects from the space
 * @author: Andias Wira-Alam <andias.alam@stud.uni-due.de>
 */

import net.jini.space.JavaSpace;
import net.jini.core.lease.Lease;

public class CleanSpace implements Runnable {
  public void run() {
    try {
      
      System.out.println("Searching for JavaSpace...");
      Lookup finder = new Lookup(JavaSpace.class);
      JavaSpace space = (JavaSpace) finder.getService();
      System.out.println("A JavaSpace has been found!\nCleaning objects from the space... ");
      int i = 0;
      while(space.readIfExists(null, null, Long.MAX_VALUE)!=null) {
        space.take(null, null, Long.MAX_VALUE);
	i++;
      }
      if(i==1) {
        System.out.print(i+ " object ");
      }
      else if(i==0) {
        System.out.print("No objects ");
      }
      else {
        System.out.print(i+ " objects ");
      }
      System.out.println("had been removed!");

      System.out.println("Done!");
    } catch(Exception e) {
        e.printStackTrace();
      }
  }
  
  public static void main(String argv[]) {
    (new Thread(new CleanSpace())).start();
  }

}
\end{lstlisting}
\begin{lstlisting}[caption=DataEntry.java]
/**
 * DataEntry.java - representing individu
 * @author: Andias Wira-Alam <andias.alam@stud.uni-due.de>
 */

import net.jini.core.entry.*;

public class DataEntry implements Entry { 
  public Integer data_id = null;

  public Integer mutationIndex = null;

  public DataEntry() {}

  public DataEntry(int id) {
    data_id = new Integer(id);
    mutationIndex = new Integer(0);
  }

  public String getDataID() {
    return "DataID: " + data_id;
  }

  public void incrementIndex() {
    mutationIndex = new Integer(mutationIndex.intValue() + 1);
  }

}
\end{lstlisting}
\begin{lstlisting}[caption=GenRandomTime.java]
/** 
 * GenRandomTime.java - generating random time for computation
 * @author: Andias Wira-Alam <andias.alam@stud.uni-due.de> 
 */ 

import java.util.*;

public class GenRandomTime { 
  
  private int low;
  
  private int up;

  private int par; // degree of parallelism
  
  private int mean;

  private int sdev; // standard deviation

  private InitValue init = new InitValue();

  public GenRandomTime() { // no-arg constructor assigns default values
    low = 100;
    up = 100;
    par = 1;
  }

  public int random() { // normal distribution random generator
    float v1, v2, s, y;
    Random gen = new Random();
    mean = (up + low) / 2;
    sdev = (up - low);
    do {
         v1 = 2 * (float) gen.nextDouble() - 1; // between -1.0 and 1.0
         v2 = 2 * (float) gen.nextDouble() - 1; // between -1.0 and 1.0
         s = v1 * v1 + v2 * v2;
    } while (s >= 1);
    float multiplier = (float) Math.sqrt(-2 * Math.log(s)/s);
    y = v1 * multiplier;
    return (int) (mean + y * sdev);
  }

  public int getTime(int step) {
    if(step==1) { // Mutation Preparation - 3 until 4 hours
      low = init.step1_low;
      up = init.step1_up;
    }
    if(step==2) { // MD Simulation - 3 until 4 days (not more than two processors)
      low = init.step2_low;
      up = init.step2_up;
    }
    if(step==3) { // Analysis - 8 until 10 hours
      low = init.step3_low;
      up = init.step3_up;
    }
    return random();
  }

  public int getPar(int step) {
    if(step==1) { 
      par = init.step1_par;
    }
    if(step==2) { 
      par = init.step2_par;
    }
    if(step==3) { 
      par = init.step3_par;
    }
    return par;
  }

}
\end{lstlisting}
\begin{lstlisting}[caption=InitValue.java]
/**
 * InitValue.java - initializing all values
 * @author: Andias Wira-Alam <andias.alam@stud.uni-due.de>
 */

public class InitValue { 
  
  public int numOfData = 8;
  
  public int numOfProcessor = 76;
  
  public int numOfStep = 3;
  
  public int numOfMutation = 10;

  /** Values for GenRandomTime */
  
  // Time for Step1: Mutation Preparation (3 until 4 hours)
  public int step1_low = 10800;  
  public int step1_up  = 14400; 
  public int step1_par = 10; // degree of parallelism
  
  // Time for Step2: MD Simulation - 3 until 4 days
  public int step2_low = 259200;  
  public int step2_up  = 345600;
  public int step2_par = 10; // degree of parallelism

  // Time for Step3: Analysis - 8 until 10 hours
  public int step3_low = 28800;
  public int step3_up  = 36000; 
  public int step3_par = 2; // degree of parallelism

  /** my no-arg constructor */
  public InitValue() {}

}
\end{lstlisting}
\begin{lstlisting}[caption=InsertEntry.java]
/**
 * InsertEntry.java - inserting all entries into the space
 * @author: Andias Wira-Alam <andias.alam@stud.uni-due.de>
 */

import net.jini.space.JavaSpace;
import java.rmi.*;
import net.jini.core.event.*;
import net.jini.core.transaction.*;
import net.jini.core.lease.*;

public class InsertEntry implements Runnable {
  public void run() {
    try {
      
      ProcessorEntry processor = new ProcessorEntry();
      DataEntry data = new DataEntry();
      InitValue init = new InitValue();
      TurnEntry turn = new TurnEntry();
      
      System.out.println("Searching for JavaSpace...");
      Lookup finder = new Lookup(JavaSpace.class);
      JavaSpace space = (JavaSpace) finder.getService();
      System.out.println("A JavaSpace has been found!");

      System.out.println("Writing objects into the space...");
      
      if(space.readIfExists(processor, null, Long.MAX_VALUE) == null)
        for(int i=1; i<=init.numOfProcessor; i++) {
	  space.write(new ProcessorEntry(i,System.currentTimeMillis()), null,
          Lease.FOREVER);
	}
      else
        System.out.println("Object \"processor\" has already existed!");

      if(space.readIfExists(data, null, Long.MAX_VALUE) == null)
        for(int i=1; i<=init.numOfData; i++) {
	  space.write(new DataEntry(i), null, Lease.FOREVER);
	}
      else
        System.out.println("Object \"data\" has already existed!");
      
      if(space.readIfExists(turn, null, Long.MAX_VALUE) == null)
        space.write(turn, null, Lease.FOREVER);
      else
        System.out.println("Object \"turn\" has already existed!");

    } catch(Exception e) {
        e.printStackTrace();
      }
  }
  
  public static void main(String argv[]) {
    try {
      Thread cleanspace = new Thread(new CleanSpace());
      Thread insertentry = new Thread(new InsertEntry());
    
      cleanspace.start(); // start the first Thread
      cleanspace.join(); // pause the execution of the next Thread until 
                         // the first finished 
      insertentry.start(); // start the second Thread
    
    } catch(Exception e) {
        e.printStackTrace();
      }
  }

}
\end{lstlisting}
\begin{lstlisting}[caption=Lookup.java]
/**
 * Lookup.java - external code by Dan Creswell
 */

import java.io.IOException;

import java.rmi.RemoteException;

import net.jini.core.lookup.ServiceRegistrar;
import net.jini.core.lookup.ServiceTemplate;

import net.jini.discovery.LookupDiscovery;
import net.jini.discovery.DiscoveryListener;
import net.jini.discovery.DiscoveryEvent;

/**
   A class which supports a simple JINI multicast lookup.  It doesn't register
   with any ServiceRegistrars it simply interrogates each one that's
   discovered for a ServiceItem associated with the passed interface class.
   i.e. The service needs to already have registered because we won't notice
   new arrivals. [ServiceRegistrar is the interface implemented by JINI
   lookup services].

   @todo Be more dynamic in our lookups - see above

   @author  Dan Creswell (dan@dancres.org)
   @version 1.00, 7/9/2003
 */
class Lookup implements DiscoveryListener {
    private ServiceTemplate theTemplate;
    private LookupDiscovery theDiscoverer;

    private Object theProxy;

    /**
       @param aServiceInterface the class of the type of service you are
       looking for.  Class is usually an interface class.
     */
    Lookup(Class aServiceInterface) {
        Class[] myServiceTypes = new Class[] {aServiceInterface};
        theTemplate = new ServiceTemplate(null, myServiceTypes, null);
    }

    /**
       Having created a Lookup (which means it now knows what type of service
       you require), invoke this method to attempt to locate a service
       of that type.  The result should be cast to the interface of the
       service you originally specified to the constructor.

       @return proxy for the service type you requested - could be an rmi
       stub or an intelligent proxy.
     */
    Object getService() {
        synchronized(this) {
            if (theDiscoverer == null) {

                try {
                    theDiscoverer =
                        new LookupDiscovery(LookupDiscovery.ALL_GROUPS);
                    theDiscoverer.addDiscoveryListener(this);
                } catch (IOException anIOE) {
                    System.err.println("Failed to init lookup");
                    anIOE.printStackTrace(System.err);
                }
            }
        }

        return waitForProxy();
    }

    /**
       Location of a service causes the creation of some threads.  Call this
       method to shut those threads down either before exiting or after a
       proxy has been returned from getService().
     */
    void terminate() {
        synchronized(this) {
            if (theDiscoverer != null)
                theDiscoverer.terminate();
        }
    }

    /**
       Caller of getService ends up here, blocked until we find a proxy.

       @return the newly downloaded proxy
     */
    private Object waitForProxy() {
        synchronized(this) {
            while (theProxy == null) {

                try {
                    wait();
                } catch (InterruptedException anIE) {
                }
            }

            return theProxy;
        }
    }

    /**
       Invoked to inform a blocked client waiting in waitForProxy that
       one is now available.

       @param aProxy the newly downloaded proxy
     */
    private void signalGotProxy(Object aProxy) {
        synchronized(this) {
            if (theProxy == null) {
                theProxy = aProxy;
                notify();
            }
        }
    }

    /**
       Everytime a new ServiceRegistrar is found, we will be called back on
       this interface with a reference to it.  We then ask it for a service
       instance of the type specified in our constructor.
     */
    public void discovered(DiscoveryEvent anEvent) {
        synchronized(this) {
            if (theProxy != null)
                return;
        }

        ServiceRegistrar[] myRegs = anEvent.getRegistrars();

        for (int i = 0; i < myRegs.length; i++) {
            ServiceRegistrar myReg = myRegs[i];

            Object myProxy = null;

            try {
                myProxy = myReg.lookup(theTemplate);

                if (myProxy != null) {
                    signalGotProxy(myProxy);
                    break;
                }
            } catch (RemoteException anRE) {
                System.err.println("ServiceRegistrar barfed");
                anRE.printStackTrace(System.err);
            }
        }
    }

    /**
       When a ServiceRegistrar "disappears" due to network partition etc.
       we will be advised via a call to this method - as we only care about
       new ServiceRegistrars, we do nothing here.
     */
    public void discarded(DiscoveryEvent anEvent) {
    }
}
\end{lstlisting}
\begin{lstlisting}[caption=ProcessorEntry.java]
/**
 * ProcessorEntry.java - representing processor
 * @author: Andias Wira-Alam <andias.alam@stud.uni-due.de>
 */

import net.jini.core.entry.*;
import java.util.ArrayList;

public class ProcessorEntry implements Entry {
  
  public Long start_time = null;

  public ArrayList<Long> recordlist = null;

  public ArrayList<ArrayList<Long>> record = null;

  public Integer proc_id = null;
  
  public ProcessorEntry() {}

  public ProcessorEntry(int id, long time) {
    start_time = new Long(time);
    proc_id = new Integer(id);
    record = new ArrayList<ArrayList<Long>>();
  }

  public void init(int id, long time) {
    start_time = new Long(time);
    proc_id = new Integer(id);
    record = new ArrayList<ArrayList<Long>>();
  }
  
  public void addRecord(long start, long stop, long data_id, long mutation, long step) {
    this.recordlist = new ArrayList<Long>();
    this.recordlist.add(new Long(start));
    this.recordlist.add(new Long(stop));
    this.recordlist.add(new Long(data_id));
    this.recordlist.add(new Long(mutation));
    this.recordlist.add(new Long(step));
    this.record.add(recordlist);
  }
  
  public long get_record(int i, int j) {
    return record.get(i).get(j);
  }

  public int get_size() {
    return record.size();
  }

  public void addRecord(long start, long stop) {
    recordlist = new ArrayList<Long>();
    recordlist.add(new Long(start));
    recordlist.add(new Long(stop));
    record.add(recordlist);
  }
  
}
\end{lstlisting}
\begin{lstlisting}[caption=RunAgentGenBased.java]
/**
 * RunAgentGenBased.java - Agent running Generation-based Algorithm
 * @author: Andias Wira-Alam <andias.alam@stud.uni-due.de>
 */

import net.jini.space.JavaSpace;
import java.rmi.*;
import net.jini.core.event.*;
import net.jini.core.transaction.*;
import net.jini.core.lease.*;

public class RunAgentGenBased implements Runnable {

  private JavaSpace space;

  public RunAgentGenBased() {
    System.out.println("Searching for JavaSpace...");
    Lookup finder = new Lookup(JavaSpace.class);
    JavaSpace space = (JavaSpace) finder.getService();
    System.out.println("A JavaSpace has been found!");
  }

  public RunAgentGenBased(JavaSpace space) {
    this.space = space;
  }

  public void run() {
    try {
      
      InitValue init = new InitValue();

      Thread[] agent = new Thread[init.numOfData];
      
      // create Threads acc. to the # of Data
      for(int i=0; i<init.numOfData; i++) { 
	  agent[i] = new Thread(new Agent(space));
      }
      
      // start Threads acc. to the # of Data
      for(int j=0; j<init.numOfData; j++) { 
          agent[j].start();
      }

      // implement Gen-based Algorithm 
      for(int k=0; k<init.numOfData; k++) { 
          agent[k].join();
      }

    } catch(Exception e) {
        e.printStackTrace();
      }
  }

  public static void main(String argv[]) {
    (new Thread(new RunAgentGenBased())).start();
  }

}
\end{lstlisting}
\begin{lstlisting}[caption=RunAgent.java]
/**
 * RunAgent.java - daemon for Agent
 * @author: Andias Wira-Alam <andias.alam@stud.uni-due.de>
 */

import net.jini.space.JavaSpace;
import java.rmi.*;
import net.jini.core.event.*;
import net.jini.core.transaction.*;
import net.jini.core.lease.*;

public class RunAgent {
  
  public static void stop() {
    System.out.println("Usage: RunAgent {gen-based|steady-state}");
    System.exit(1);
  }

  public static void main(String argv[]) {
    try {
      
      InitValue init = new InitValue();
      Lookup finder = null;
      JavaSpace space = null;

      Thread[] agent = null;
      
      if(argv.length==0 || argv.length>=2) {
        stop();
      }
      
      if(argv[0].equals("gen-based") || argv[0].equals("steady-state")) {
        System.out.println("Searching for JavaSpace...");
        finder = new Lookup(JavaSpace.class);
        space = (JavaSpace) finder.getService();
        System.out.println("A JavaSpace has been found!");
      }
      
      if(argv[0].equals("gen-based")) {
        agent = new Thread[init.numOfMutation];
        for(int i=0; i<init.numOfMutation; i++) { // Threads for Gen-based Algorithm
          agent[i] = new Thread(new RunAgentGenBased(space));
        }
        for(int j=0; j<init.numOfMutation; j++) { // implement Gen-based Algorithm 
          agent[j].start();
          agent[j].join();
        }
      }
      else if(argv[0].equals("steady-state")) {
        agent = new Thread[init.numOfData];
        for(int i=0; i<init.numOfData; i++) { // Threads for Steady-state Algorithm
          agent[i] = new Thread(new RunAgentSteadyState(space));
        }
	for(int j=0; j<init.numOfData; j++) { // implement Steady-state Algorithm
          agent[j].start();
        }
      }
      else 
        stop();

    } catch(Exception e) {
        e.printStackTrace();
      }
  }
}
\end{lstlisting}
\begin{lstlisting}[caption=RunAgentSteadyState.java]
/**
 * RunAgentSteadyState.java - Agent running Steady State Algorithm
 * @author: Andias Wira-Alam <andias.alam@stud.uni-due.de>
 */

import net.jini.space.JavaSpace;
import java.rmi.*;
import net.jini.core.event.*;
import net.jini.core.transaction.*;
import net.jini.core.lease.*;

public class RunAgentSteadyState implements Runnable {
  
  private JavaSpace space;

  public RunAgentSteadyState(JavaSpace space) {
    this.space = space;
  }

  public RunAgentSteadyState() {
    System.out.println("Searching for JavaSpace...");
    Lookup finder = new Lookup(JavaSpace.class);
    JavaSpace space = (JavaSpace) finder.getService();
    System.out.println("A JavaSpace has been found!");
  }

  public void run() {
    try {
      
      InitValue init = new InitValue();
      Thread[] agent = new Thread[init.numOfMutation];

      for(int i=0; i<init.numOfMutation; i++) { // iterate according to the number of Mutation
          agent[i] = new Thread(new Agent(space));
      } 
      for(int j=0; j<init.numOfMutation; j++) { // iterate according to the number of Mutation
          agent[j].start();
          agent[j].join();
      } 

    } catch(Exception e) {
        e.printStackTrace();
      }
  }
  
  public static void main(String argv[]) {
    (new Thread(new RunAgentSteadyState())).start();
  }

}
\end{lstlisting}
\begin{lstlisting}[caption=TakeProcessor.java]
/**
 * TakeProcessor.java - take processor objects from the space
 * @author: Andias Wira-Alam <andias.alam@stud.uni-due.de>
 */

import net.jini.space.JavaSpace;
import net.jini.core.lease.Lease;

public class TakeProcessor {
  public static void main(String argv[]) {
    try {
      
      System.out.println("Searching for JavaSpace...");
      Lookup finder = new Lookup(JavaSpace.class);
      JavaSpace space = (JavaSpace) finder.getService();
      System.out.println("A JavaSpace has been found!");
      
      ProcessorEntry ptemplate = new ProcessorEntry();
      ProcessorEntry presult = new ProcessorEntry();
      InitValue init = new InitValue();
      System.out.println("Taking processor objects from the space... ");
      for(int i=1; i <= init.numOfProcessor; i++) {
        presult = (ProcessorEntry) space.takeIfExists(ptemplate, null, Long.MAX_VALUE);
        if(presult!=null) {
	  for(int j=0; j<presult.record.size(); j++) {
	    System.out.print("ProcID: " +presult.proc_id+ "; ");
	    System.out.print("start_t: " +presult.record.get(j).get(0)+ "; ");
	    System.out.print("stop_t: " +presult.record.get(j).get(1)+ "; ");
	    System.out.print("data_id: " +presult.record.get(j).get(2)+ "; ");
	    System.out.print("mutation: " +presult.record.get(j).get(3)+ "; ");
	    System.out.println("step: " +presult.record.get(j).get(4));
	  }
	}
	else {
          System.out.println("Processor Objects NOT FOUND!");
	  break;
	}  
      }
      System.out.println("Done!");
    } catch(Exception e) {
        e.printStackTrace();
      }
  }
}
\end{lstlisting}
\begin{lstlisting}[caption=TestSpace.java]
/**
 * TestSpace.java - checking objects availability in the space
 * @author: Andias Wira-Alam <andias.alam@stud.uni-due.de>
 */

import net.jini.space.JavaSpace;
import net.jini.core.lease.Lease;

public class TestSpace {
  public static void main(String argv[]) {
    try {
      
      System.out.println("Searching for JavaSpace...");
      Lookup finder = new Lookup(JavaSpace.class);
      JavaSpace space = (JavaSpace) finder.getService();
      System.out.println("A JavaSpace has been found!");
      
      ProcessorEntry ptemplate = new ProcessorEntry();
      DataEntry dtemplate = new DataEntry();

      if(space.readIfExists(ptemplate, null, Long.MAX_VALUE) != null)
          System.out.println("Object \"processor\" has been found!");
      else    
	  System.out.println("Object \"processor\" doesn't exist!");
      
      if(space.readIfExists(dtemplate, null, Long.MAX_VALUE) != null)
          System.out.println("Object \"data\" has been found!");
      else    
	  System.out.println("Object \"data\" doesn't exist!");

    } catch(Exception e) {
        e.printStackTrace();
      }
  }
}
\end{lstlisting}
\begin{lstlisting}[caption=TurnEntry.java]
/**
 * TurnEntry.java - performing "turn token"
 * @author: Andias Wira-Alam <andias.alam@stud.uni-due.de>
 */

import net.jini.core.entry.*;

public class TurnEntry implements Entry {
  public String name = null;
  public Integer value = null;
  
  public TurnEntry() {}
}
\end{lstlisting}
\end{scriptsize}

\chapter{Figures: Histogram}\label{appendixB}

This appendix contains the histograms of the overall computation time
of all scenarios.

\begin{figure}[htp]
\centering
\includegraphics[scale=1]{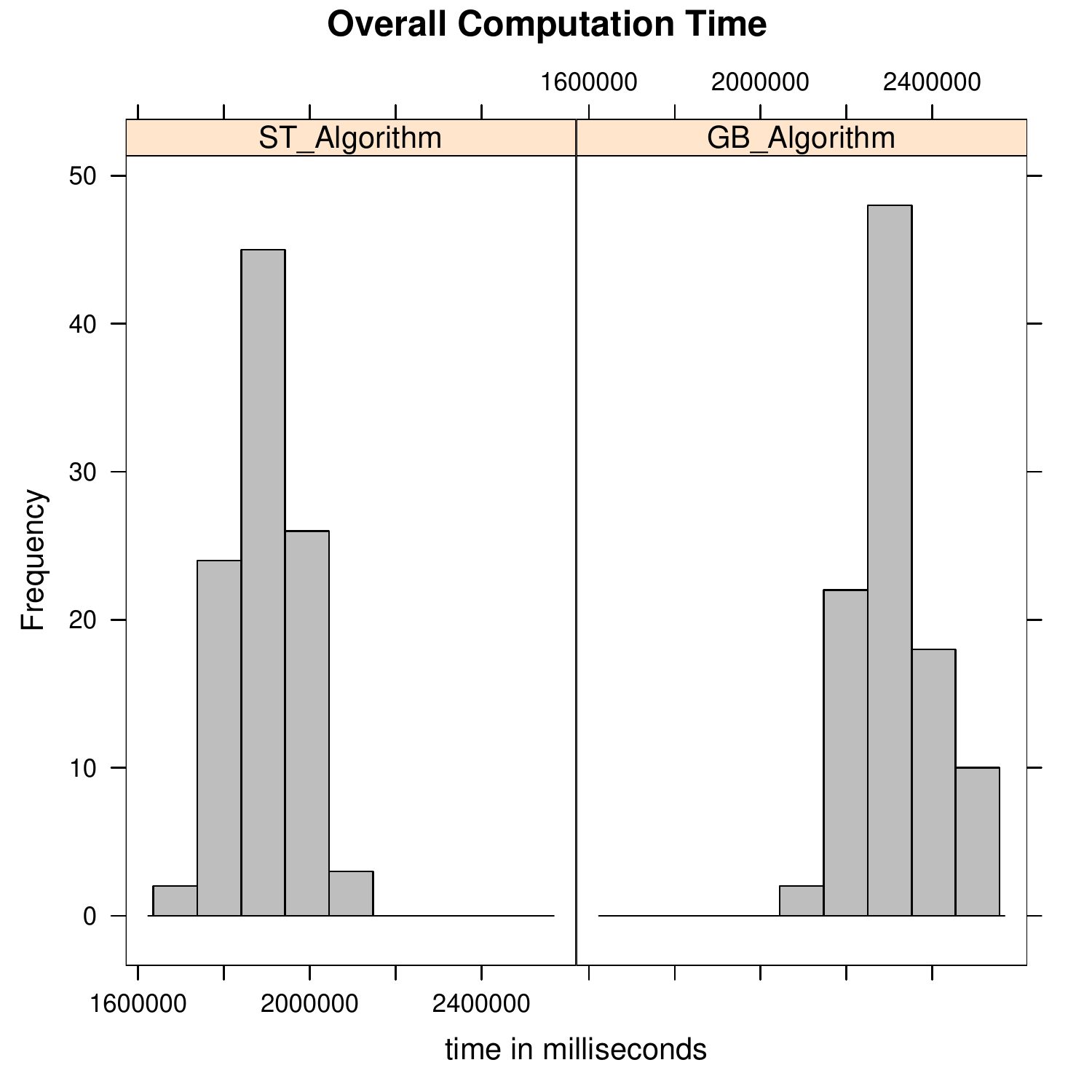}
\caption{Histogram of scenario e00}
\label{hist_e00}
\end{figure}

\begin{figure}[htp]
\centering
\includegraphics[scale=1]{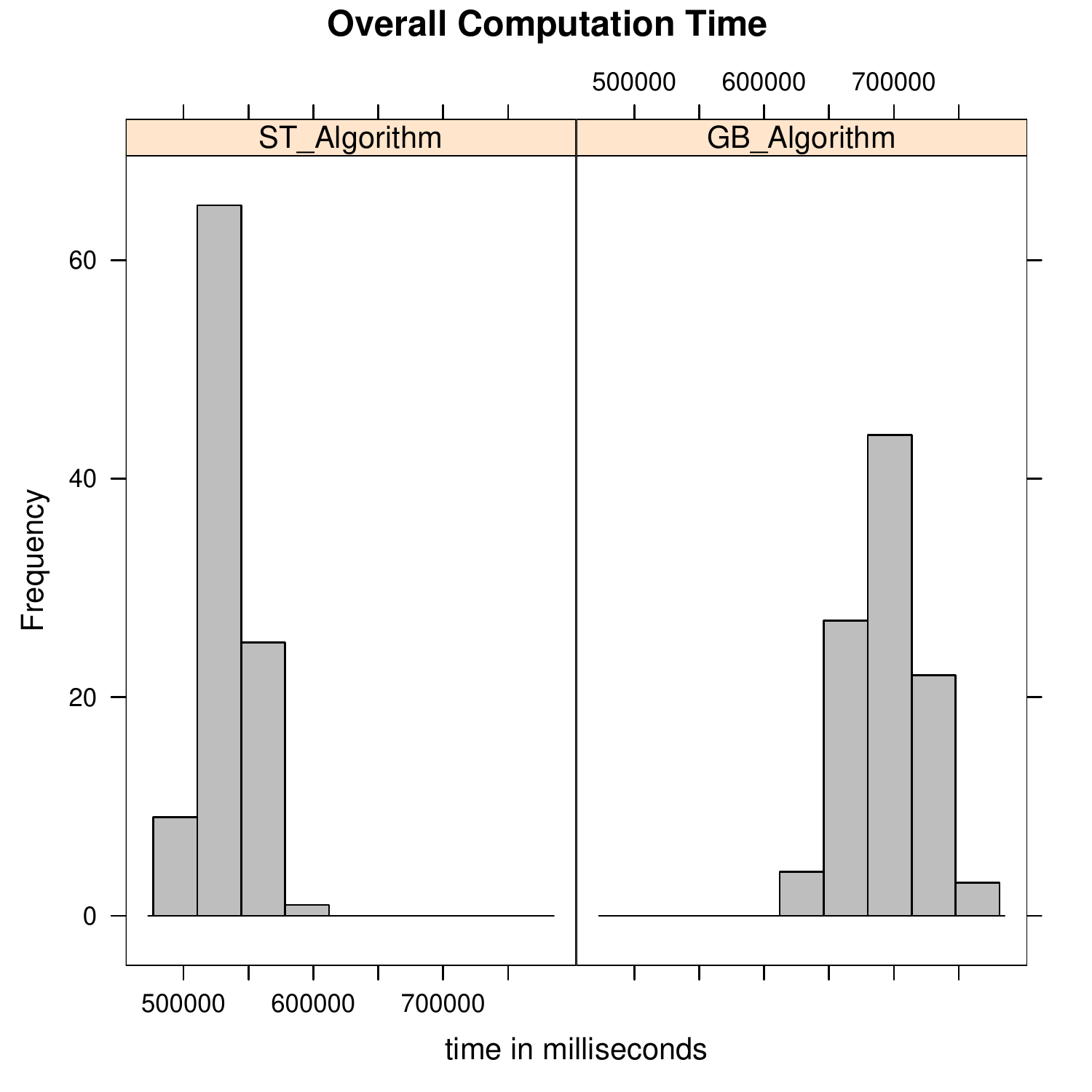}
\caption{Histogram of scenario e02}
\label{hist_e02}
\end{figure}

\begin{figure}[htp]
\centering
\includegraphics[scale=1]{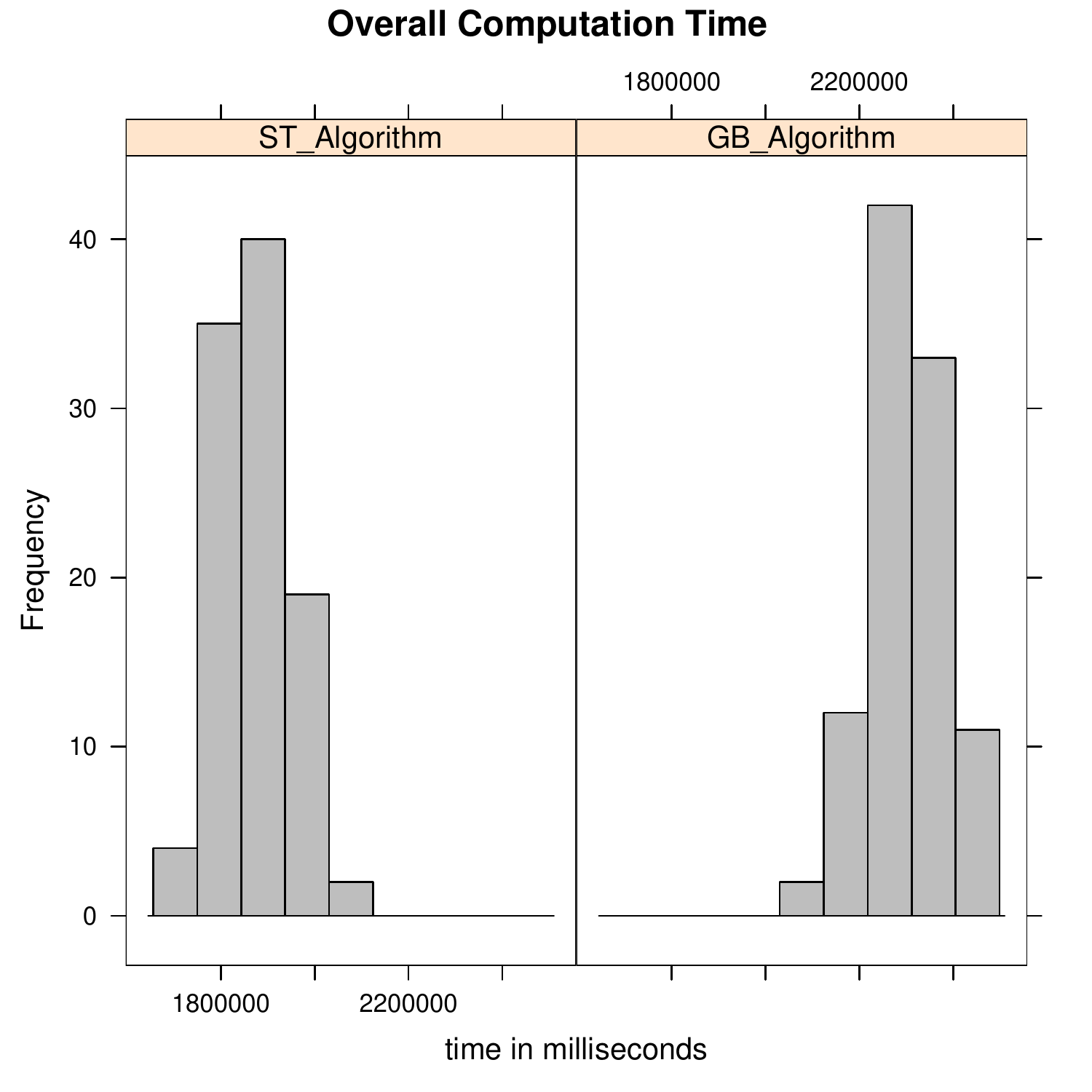}
\caption{Histogram of scenario e12}
\label{hist_e12}
\end{figure}

\begin{figure}[htp]
\centering
\includegraphics[scale=1]{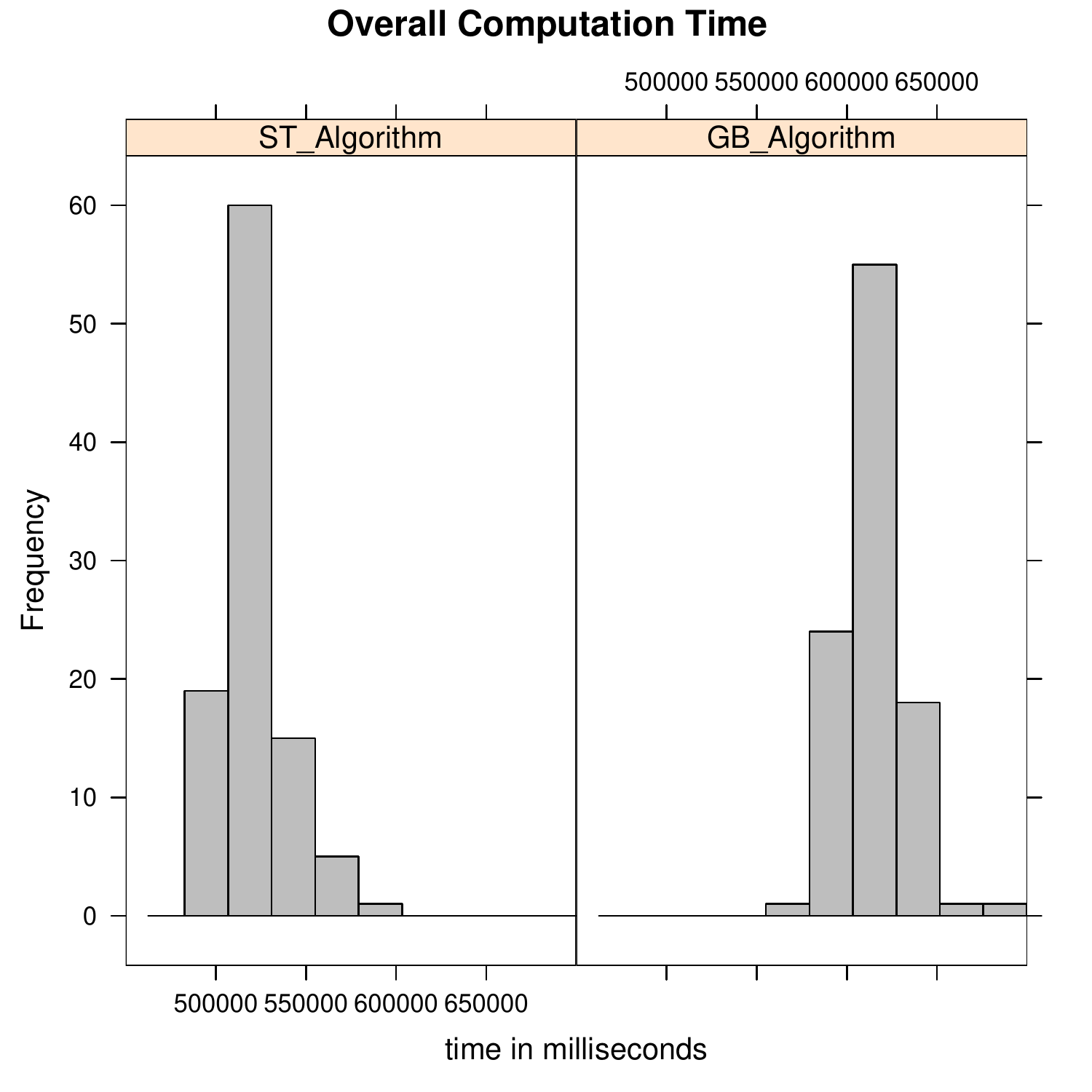}
\caption{Histogram of scenario e14}
\label{hist_e14}
\end{figure}

\end{appendix}

\bibliographystyle{plain}  

\bibliography{WiraAlam_Masterthesis}
\nocite{*}

\end{document}